\documentclass[twocolumn,pre,aps,showpacs]{revtex4}

\usepackage{graphicx}
\usepackage{bbold}
\usepackage{amsmath}
\usepackage{float}
\usepackage{color}

\def\be{\begin{eqnarray}}
\def\ee{\end{eqnarray}}
\def\ben{\begin{eqnarray*}}
\def\een{\end{eqnarray*}}
\def\bes{\begin{subequations}}
\def\ees{\end{subequations}}
\def\ds{\displaystyle}
\def\nn{\nonumber}
\def\erfc{\rm erfc}

\begin{document}

\title{Accurate Determination of the Shear Viscosity of the One-Component Plasma}

\author{J\'er\^ome \surname{Daligault}}\email{daligaul@lanl.gov}
\affiliation{Theoretical Division, Los Alamos National Laboratory, Los Alamos, NM 87545, USA}
\author{Kim \O.~Rasmussen}
\affiliation{Theoretical Division, Los Alamos National Laboratory, Los Alamos, NM 87545, USA}
\author{Scott D.\ Baalrud}
\affiliation{Department of Physics and Astronomy, University of Iowa, Iowa City, Iowa 52242, USA}

\begin{abstract}
The shear viscosity coefficient of the one-component plasma is calculated with unprecedented accuracy using equilibrium molecular dynamics simulations and the Green-Kubo relation.
Numerical and statistical uncertainties and their mitigation for improving accuracy are analyzed.
In the weakly coupled regime, our the results agree with the Landau-Spitzer prediction.
In the moderately and strongly coupled regimes, our results are found in good agreement with recent results obtained for the Yukawa one-component plasma using non-equilibrium molecular dynamics.
A practical formula is provided for evaluating the viscosity coefficient across coupling regimes, from the weakly-coupled regime up to solidification threshold.
The results are used to test theoretical predictions of the viscosity
coefficients found in the literature.
\end{abstract}


\date{\today}

\maketitle

\section{Introduction}

Like the hard-sphere model in the theory of simple liquids, the classical one-component plasma (OCP) is a reference model in the study of strongly coupled Coulomb systems and, in particular, of ions in strongly coupled plasmas \cite{BausHansen1980}.
By definition, the OCP consists of a system of identical ions of charge $Ze$, mass $m$ and number density $n$ in an infinite three-dimensional space.
Particle dynamics is governed by the laws of classical, non-relativistic mechanics.
The interaction energy between two ions separated by the distance $r$ is modeled by a Yukawa potential $\ds v(r)=q^2\frac{e^{-r/\lambda_{sc}}}{r}$, where $\lambda_{sc}\geq 0$ is a parameter used to describe the screening effect of the conduction electrons on the bare ion-ion Coulomb interactions, and $q^2=(Ze)^2/4\pi\epsilon_0$.
In the limit $\lambda_{sc} \rightarrow +\infty$, particles interact via the bare Coulomb interaction and the ions must be immersed in a uniform, neutralizing background for well-posedness of the model.

The equilibrium properties of the OCP depend on only two dimensionless parameters: the screening parameter $\kappa=a/\lambda_{sc}$ and the Coulomb coupling parameter $\Gamma=q^2/ak_BT$, where $a=(3/4\pi n)^{1/3}$ is the Wigner-Seitz radius and $T$ is the temperature.
The Coulomb coupling parameter measures the degree of non-ideality of the system, i.e. the degree to which many-body interactions affect the properties of the ensemble of ions.
Given a value for $\kappa$, the OCP shows transitions from a nearly collisionless, gaseous regime for $\Gamma<<1$ continuously through an increasingly correlated, liquid-like regime to the Wigner crystallization into a lattice near $\Gamma_m$ (e.g., $\Gamma_m\simeq 175$ at $\kappa=0$, $\Gamma_m=440$ at $\kappa=2$).
The gas-like to liquid-like crossover manifests itself in several ways in the microscopic properties of the OCP.
Most noticeably, the coefficient of reduced shear viscosity
\be
\eta^*=\frac{\eta}{mna^2\omega_p}\,,
\ee
where $\omega_p$ is the plasma frequency defined below, exhibits a minimum at intermediate values around $\Gamma_{min}\sim \Gamma_m/10$ \cite{SaigoHamaguchi2002,VieillefosseHansen1975}.
In absolute units, the shear-viscosity coefficient $\eta$ increases monotonically with density along any isotherm, whereas along any isochore, $\eta$ exhibits a minimum as a function of temperature.
In a fluid, transport of momentum occurs not only by the bodily movement of particles, but also by the direct transmission of intermolecular forces, which results from a competition between kinetic and interaction effects.
At small coupling $\Gamma\ll\Gamma_{min}$, the former mechanism is predominant and, like in a gas, the OCP viscosity increases with increasing temperature.
At large coupling $\Gamma \gg \Gamma_{min}$, the latter mechanism is predominant and, like in a liquid, the OCP viscosity decreases with increasing temperature. Strong interparticle interactions give rise to the cage-effect \cite{Daligault2006}, whereby each particle finds itself trapped for some period of time in the cage formed by its immediate neighbors, rebounding until it overcomes the energy barrier and diffuses to a neighboring cage.
At intermediate coupling $\Gamma\sim\Gamma_{min}$, the two momentum transport mechanisms contribute with similar magnitude, resulting in a shallow minimum in the viscosity coefficient.

\begin{figure}[t]
\includegraphics[width=\columnwidth]{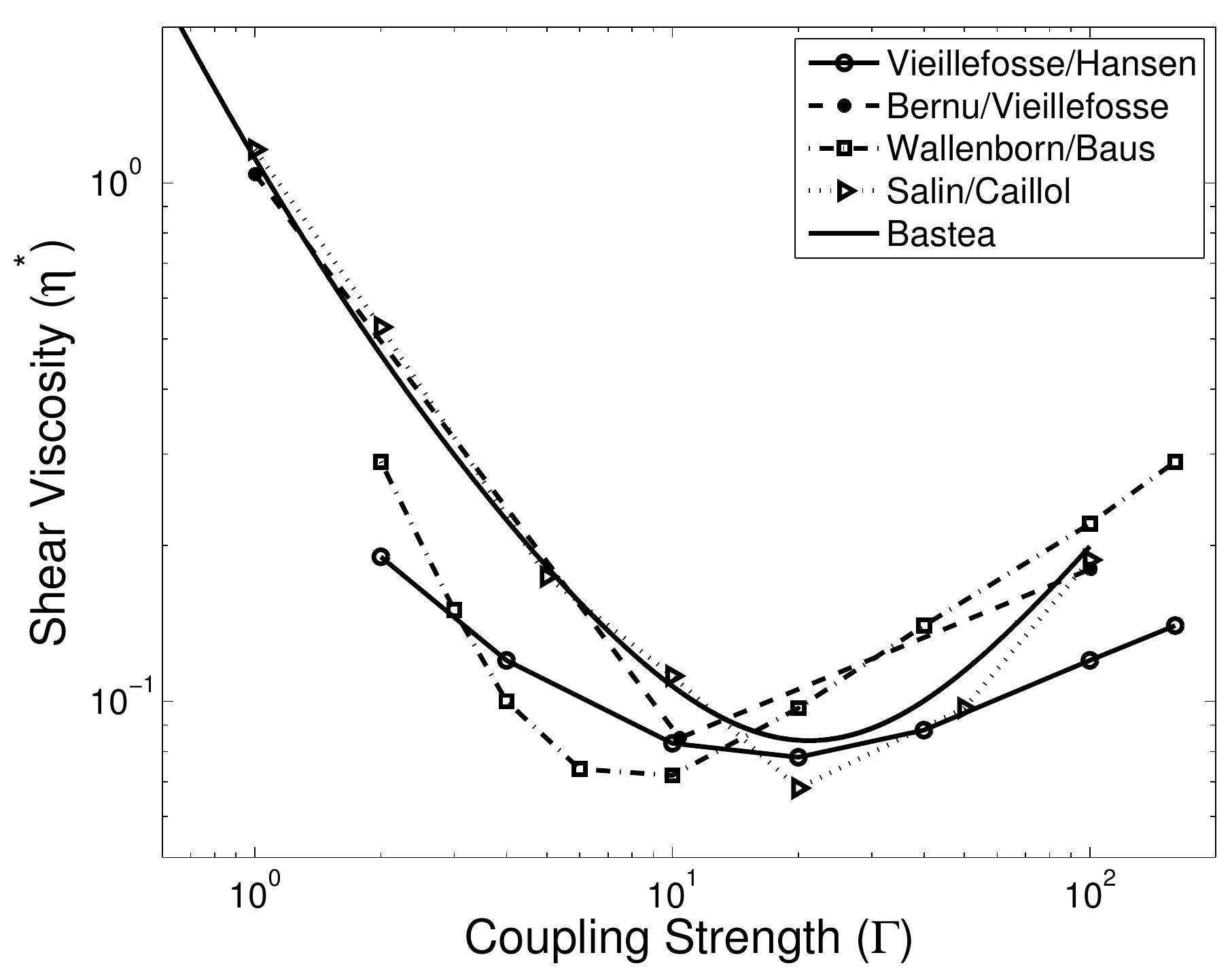}
\caption{Shear-viscosity coefficient of the Coulomb OCP ($\kappa=0$) obtained by different authors \cite{bern:78,wall:78,SalinCaillol2003,bast:05} using equilibrium molecular dynamics and the Green-Kubo relation. Results obtained using non-equilibrium molecular dynamics are compiled in \cite{DonkoHartmann2008}.}
\label{Fig0}
\end{figure}

Despite the apparent simplicity of the OCP model, accurate determination of the viscosity coefficient of the OCP by molecular dynamics (MD) simulations is difficult.
This is exemplified by the significant discrepancies in the results obtained over the years by different authors.
A compilation of these has been provided in \cite{DonkoHartmann2008}.
Remarkably, important differences are found not only between results obtained using different MD techniques, but also between those obtained using the same technique.
For example, previous results obtained using equilibrium MD are shown in Figure~\ref{Fig0}.
Among these, the results of Bastea \cite{bast:05} are believed to be most accurate.
Recently, Donk{\'o} and Hartmann have presented arguably the most accurate results for moderately and strongly coupled OCP's at $\kappa=1,2,3$ using two independent {\it non-equilibrium} MD simulation methods, namely the M{\"u}ller-Plathe reverse MD approach and the Evans-Morriss homogeneous shear algorithm.
In the present paper, we use {\it equilibrium} MD based on the evaluation of the Green-Kubo relation to validate the non-equilibrium MD results of Donk{\'o} and Hartmann and the equilibrium MD results of Bastea for the Coulomb OCP ($\kappa=0$).

There are a couple of reasons why this is important.
Donk{\'o} and Hartmann chose non-equilibrium MD methods, claiming that they are generally more efficient than equilibrium calculations. 
Indeed, we shall see that the determination of viscosity from the Green-Kubo relation is made difficult by the large statistical imprecision in the calculation of the shear-stress autocorrelation function.
This arises mainly due to the fact that simulation averages are taken over finite-length runs.
The noise can be satisfactorily reduced at the price of very long simulations, but this requires much longer run times than have previously been reported.
Despite the computational cost, equilibrium MD has advantages.
It provides information about the microphysical ion dynamics, in particular the time-correlation function of the shear stress.
It works equally well for all $\kappa$ and $\Gamma$ values, including the Coulomb OCP.
The same simulation can be used to consistently calculate all other transport and static properties of the plasma.
Furthermore,  it provides an independent method that allows us to confirm Donk{\'o} and Hartmann's results.

In the past, much effort has been devoted to develop a theory that extends the traditional plasma regime valid at small $\Gamma$ to the moderate and strongly coupled regimes.
An accurate determination of the shear-viscosity is desirable to test existing and future theories.
In this paper, we test the conventional result of Landau-Spitzer, the kinetic theories of Wallenborn-Baus~\cite{wall:78}, Viellifosse-Hansen~\cite{VieillefosseHansen1975} and of Tanaka-Ichimaru~\cite{tana:86}, and the recent effective potential theory of Baalrud-Daligault~\cite{BaalrudDaligault2013prl}.

This paper is organized as follows.
Section~\ref{section_II} describes the equilibrium simulations used to determine the shear viscosity coefficient.
We present a detailed study of the statistical convergence necessary to ensure quality of the final results.
The simulation methods and parameters are explicitly given to help anyone who wishes to reproduce our results.
The results are described in section~\ref{section_III}.
Finally, in section~\ref{section_IV}, we compare the results with the above mentioned theories.

\section{Simulation Methods and Results} \label{section_II}

In the following, $\omega_p=(4\pi n q^2/m)^{1/2}$ denotes the plasma frequency.
In our MD simulation $1/\omega_p$ and the Wigner-Seitz radius $a$ are used as unit of time and length, respectively.

\subsection{Basic definitions}

\begin{table}
\begin{tabular}{l|l}
Particle number & $5000$ ($\Gamma\geq .5$) ; $50000$ ($\Gamma<0.5$)\\
Time step & $0.01/\omega_p$; $0.001/\omega_p$ ($\Gamma<0.5$)\\
Simulation length & $83886.08/\omega_p$\\
Equilibration length & $1000/\omega_p$\\
Numerical method & Ewald sums with $\rm P^3M$ algorithm \cite{HockneyEastwood}\\
Ewald parameter $a\alpha$ & 0.64 ($N=5000$) 1 (N=50000)\\
Short-range cutoff $r_C/a$ & 5.0\\
FFT grid & $54^3$ (N=5000) $64^3$ (N=50000)\\
rms forces & $<\,10^{-5}$
\end{tabular}
\caption{Main parameters of the equilibrium MD simulations used in this work to calculate the viscosity coefficients.}
\label{table1}
\end{table}
Here we describe a typical equilibrium MD simulation of an OCP at given $\Gamma$ and $\kappa$.
$N$ particles are placed in a cubic box of volume $V=L^3$ and periodic conditions are imposed on all boundaries.
Particle trajectories are determined by solving Newton's equations of motion with the velocity Verlet integrator \cite{FrenkelSmit}.
The force on an ion that results from its interaction with the ions in the simulation box and with those in the periodically replicated cells is calculated using the Ewald summation technique.
This is essential for small $\kappa_{sc}:=a/\lambda_{sc}$ values because the range of the interaction is larger than the simulation box in this case.
A formulation of the Ewald sum approach for Yukawa potentials can be found in \cite{SalinCaillol2003}.
The interaction energy $v(r)$ between two particles at distance $r$ is represented by a sum of a short-range (sr) and a long-range (lr) component,
\be
v(r)=q^2\phi_{\rm sr}(r)+q^2\phi_{\rm lr}(r)
\ee
where
\be
\phi_{\rm sr}(r)\!=\!\!\frac{1}{2r}\left[\erfc\left(\alpha r+\frac{\kappa_{sc}}{2\alpha}\right)e^{\kappa_{sc} r}+\erfc\left(\alpha r-\frac{\kappa_{sc}}{2\alpha}\right)e^{-\kappa_{sc} r}\right]\nonumber
\ee
and
\be
\phi_{\rm lr}(r)\!=\!\!\frac{4\pi }{V}\sum_{{\bf n}\in\mathbb{Z}^3}{\frac{e^{-(k^2+\kappa_{sc}^2)/(4\alpha^2)}}{k^2+\kappa_{sc}^2} e^{i{\bf k}\cdot{\bf r}}}\quad,\quad{\bf k}=\frac{2\pi}{L}{\bf n}
\ee
where $\alpha>0$ is the Ewald parameter, and $\erfc$ is the complementary error function.
In our simulation code, the Ewald sum is calculated with the particle-particle-particle-mesh ($\rm P^3M$) method, which combines high-resolution of close encounters (the sr term is calculated using nearest neighbor techniques) and rapid, long-range force calculations (the lr forces are computed on a mesh using three-dimensional fast Fourier transforms) \cite{HockneyEastwood}.

Table \ref{table1} lists the main numerical parameters used in the present study to compute the viscosity coefficients.
Our timestep $\delta t = 10^{-2} /\omega_p$ is chosen small enough to ensure excellent energy conservation for all $\Gamma$ and $\kappa$ values.
In all simulations, $N=5000$ for $\Gamma>0.5$, while $N=50000$ was chosen for $\Gamma<0.5$ to ensure high enough collision ability in the simulation cell.

\subsection{Shear viscosity coefficient}

 The shear viscosity coefficient, $\eta$, was computed using the Green-Kubo relation that expresses $\eta$ as the time integral of the equilibrium autocorrelation function of the off-diagonal components $\sigma_{xy}$ of the shear stress tensor $\tensor\sigma$ \cite{HansenMcDonald},
\be
\eta=\frac{1}{6Vk_B T}\sum_{x=1}^3{\sum_{y\neq x=1}^3{\int_0^\infty{J_{xy}(t) dt}}}\,. \label{GK_viscosity}
\ee
where $J_{xy}(t)$ is the shear-stress autocorrelation function
\be
J_{xy}(t)=\big\langle \sigma_{xy}(t)\sigma_{xy}(0) \big\rangle_{eq}\,. \label{shearstressautocorrelationfunction}
\ee
In Eq.(\ref{GK_viscosity}), the brackets $\big\langle \dots \big\rangle_{eq}$ the denote equilibrium (thermal) average at temperature $T$.
As shown Ref. \cite{SalinCaillol2003}, it follows from the Ewald decomposition of the interaction potential that the components of the shear stress tensor can be conveniently split into a kinetic component, a short-range interaction component,  and a long-range interaction component
\be
\tensor\sigma(t)=\tensor\sigma^{\,{\rm kin}}(t)+\tensor\sigma^{\,{\rm sr}}(t)+\tensor\sigma^{\,{\rm lr}}(t)\,,\label{sigma_decomposition}
\ee
where
\begin{widetext}
\be
\tensor\sigma^{\,{\rm kin}}(t)&=&\sum_{i=1}^N{m{\bf v}_i(t){\bf v}_i(t)}\label{sigma_kin}\\
\tensor\sigma^{\,{\rm sr}}(t)&=&-\frac{q^2}{2}\sum_{i\neq j=1}^N{\sum_{{\bf n}\in\mathbb{Z}^3}{\frac{({\bf r}_{ij}(t)+{\bf n}L)({\bf r}_{ij}(t)+{\bf n}L)}{||{\bf r}_{ij}(t)+{\bf n}L||} \phi_{\rm sr}^\prime(||{\bf r}_{ij}(t)+{\bf n}L||)}} \label{sigma_sr}\\
\tensor\sigma^{\,{\rm lr}}(t)&=&\frac{q^2}{2V}\sum_{i\neq j=1}^N{\sum_{{\bf n}\in\mathbb{Z}^3}{\left(\tilde\phi_{\rm lr}(k) \stackrel{\leftrightarrow}{\rm U}+\frac{d\tilde\phi_{\rm lr}(k)}{dk}\frac{{\bf k}{\bf k}}{k} \right)e^{i{\bf k}\cdot{\bf r}_{ij}(t)}}}\quad,\quad{\bf k}=\frac{2\pi}{L}{\bf n}\label{sigma_lr}
\ee
\end{widetext}
in which ${\bf r}_i(t)$ and ${\bf v}_i(t)$ are the instantaneous postion and velocity of particle $i$ at time $t$, ${\bf r}_{ij}={\bf r}_i-{\bf r}_j$, $\stackrel{\leftrightarrow}{\rm U}$ is the unit dyad tensor, and $\tilde\phi_{\rm lr}(k)=4\pi\frac{e^{-(k^2+\kappa_{sc}^2)/(4\alpha^2)}}{k^2+\kappa_{sc}^2}$.

\subsection{Typical MD simulation}

\begin{figure*}[htb]
\includegraphics[scale=.4]{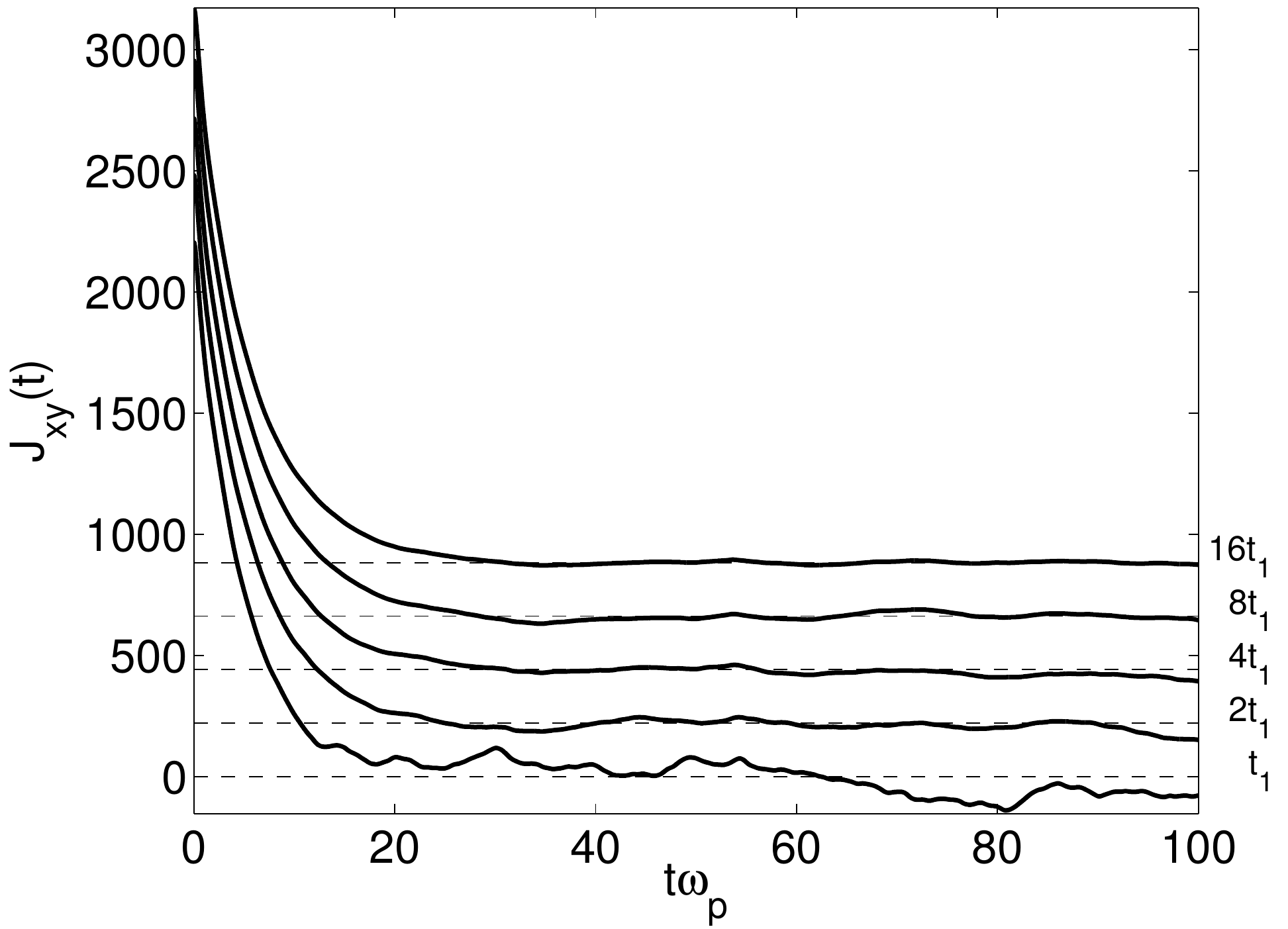}
\includegraphics[scale=.4]{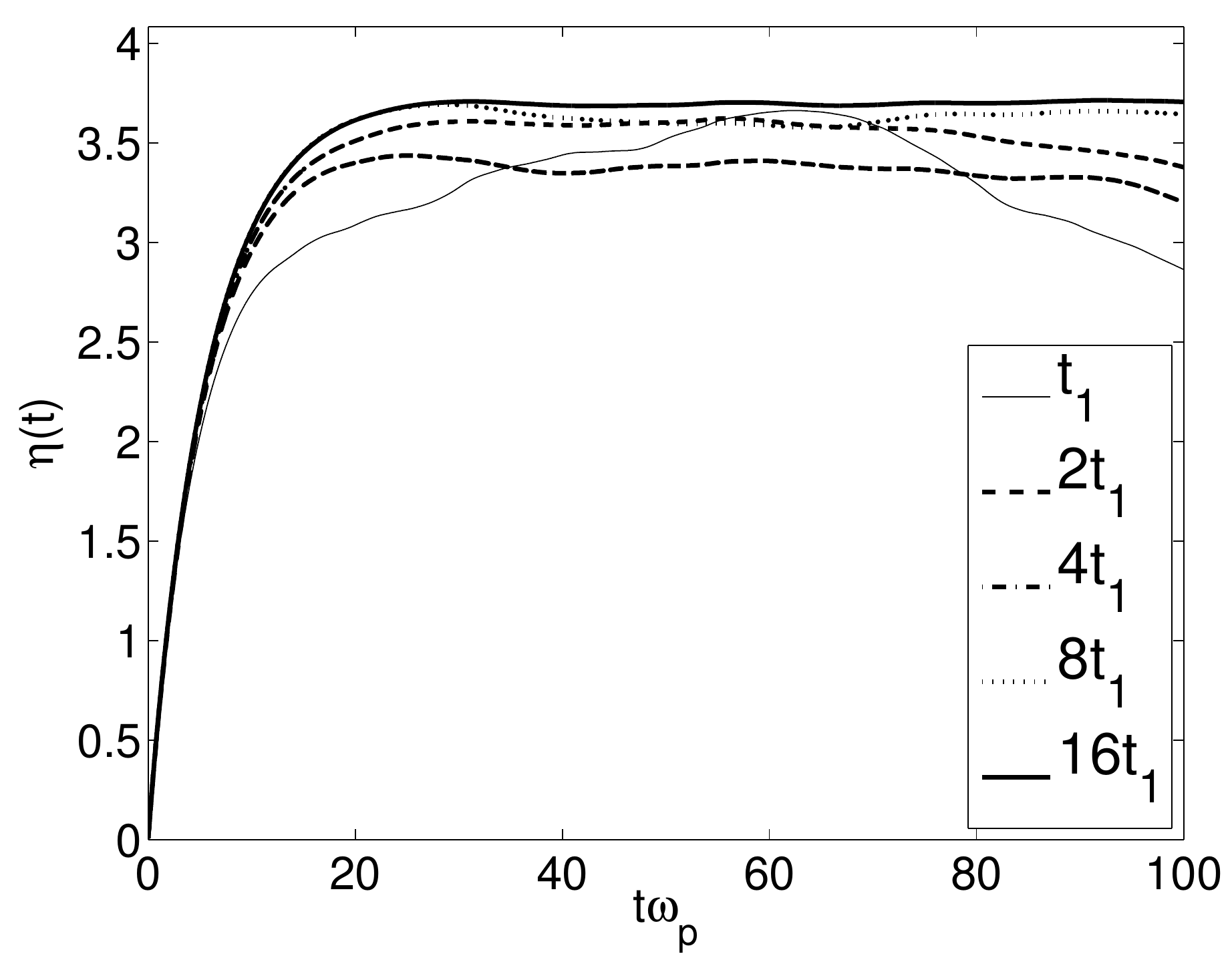}\\
\includegraphics[scale=.4]{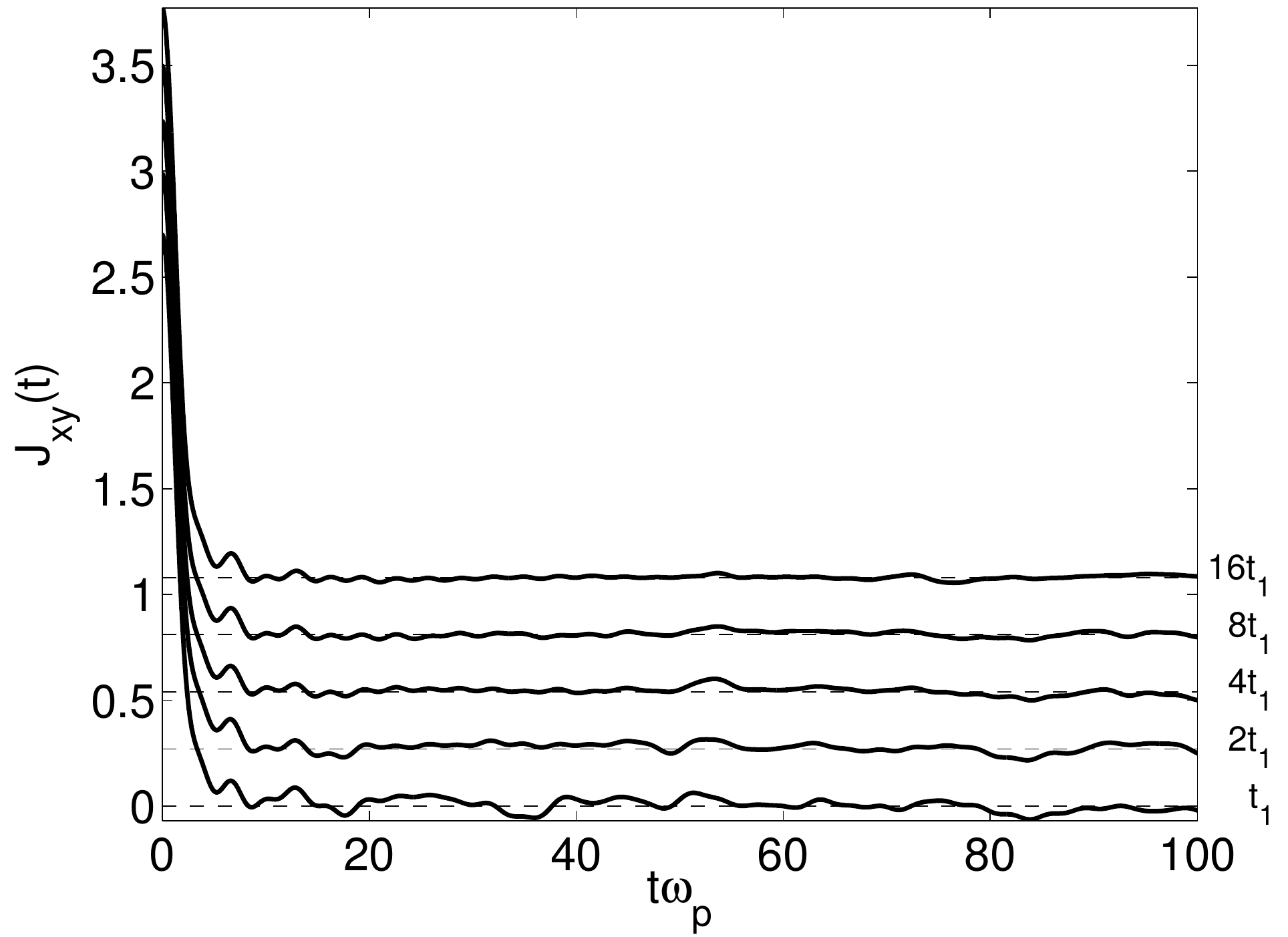}
\includegraphics[scale=.4]{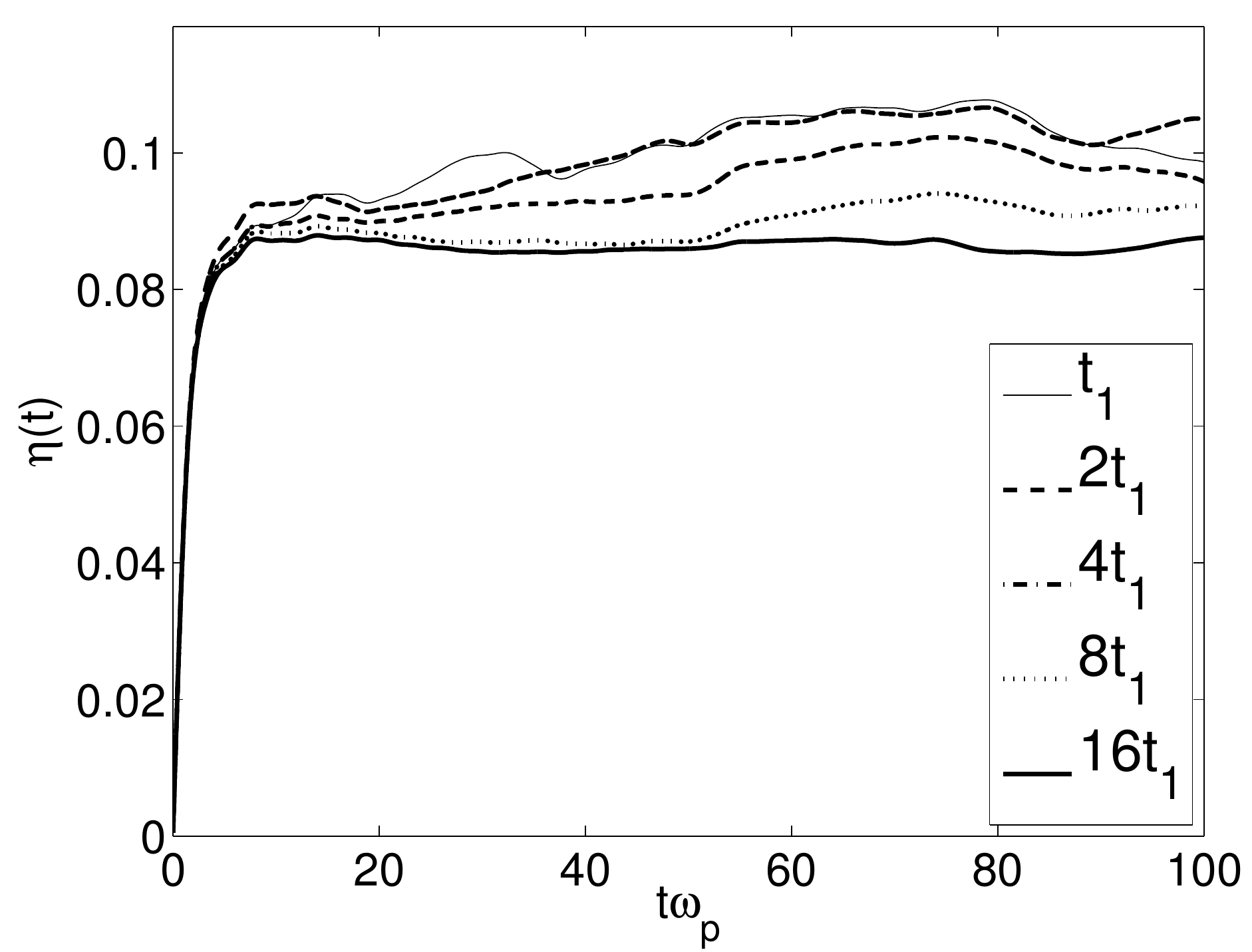}\\
\includegraphics[scale=.4]{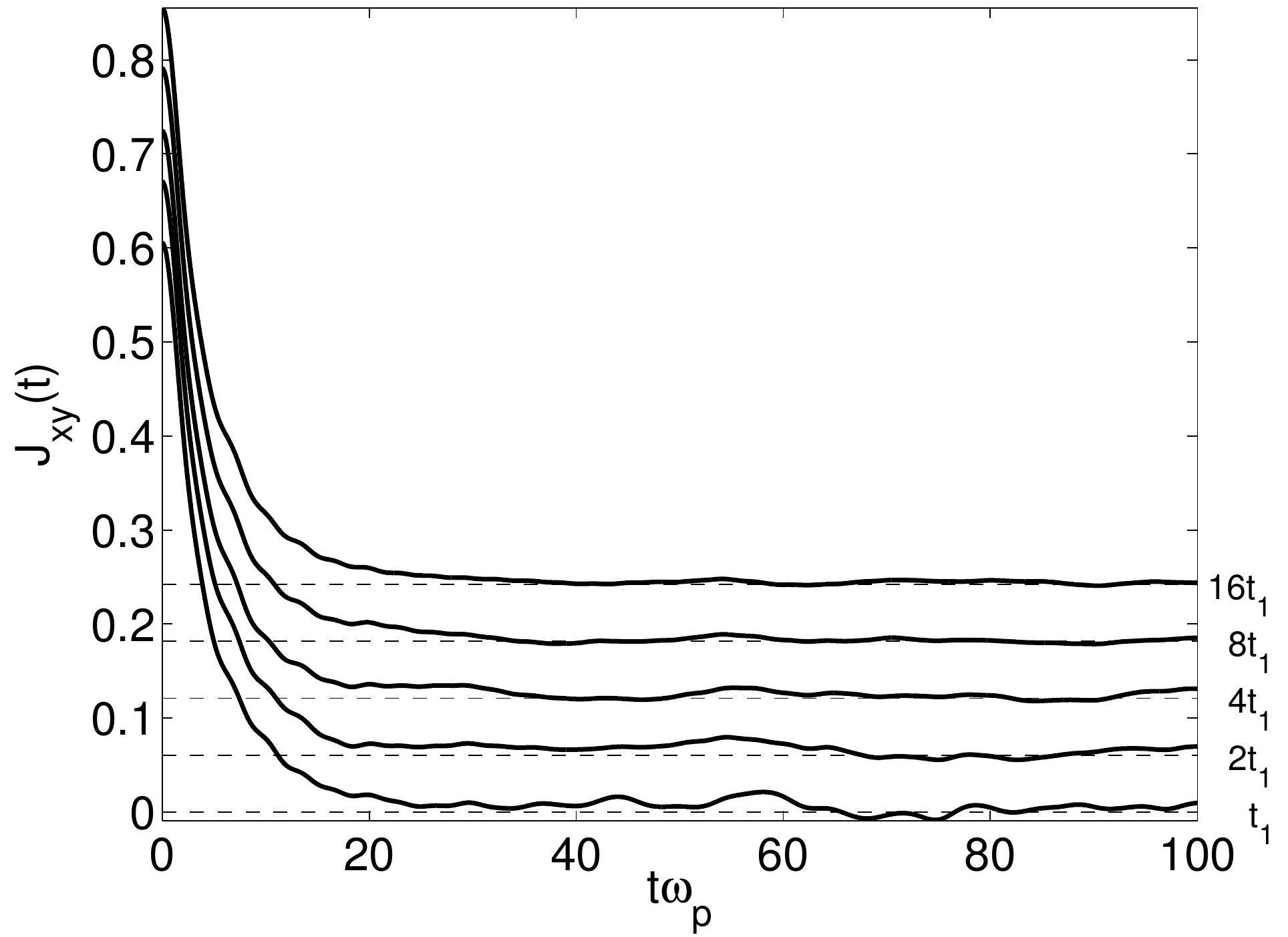}
\includegraphics[scale=.4]{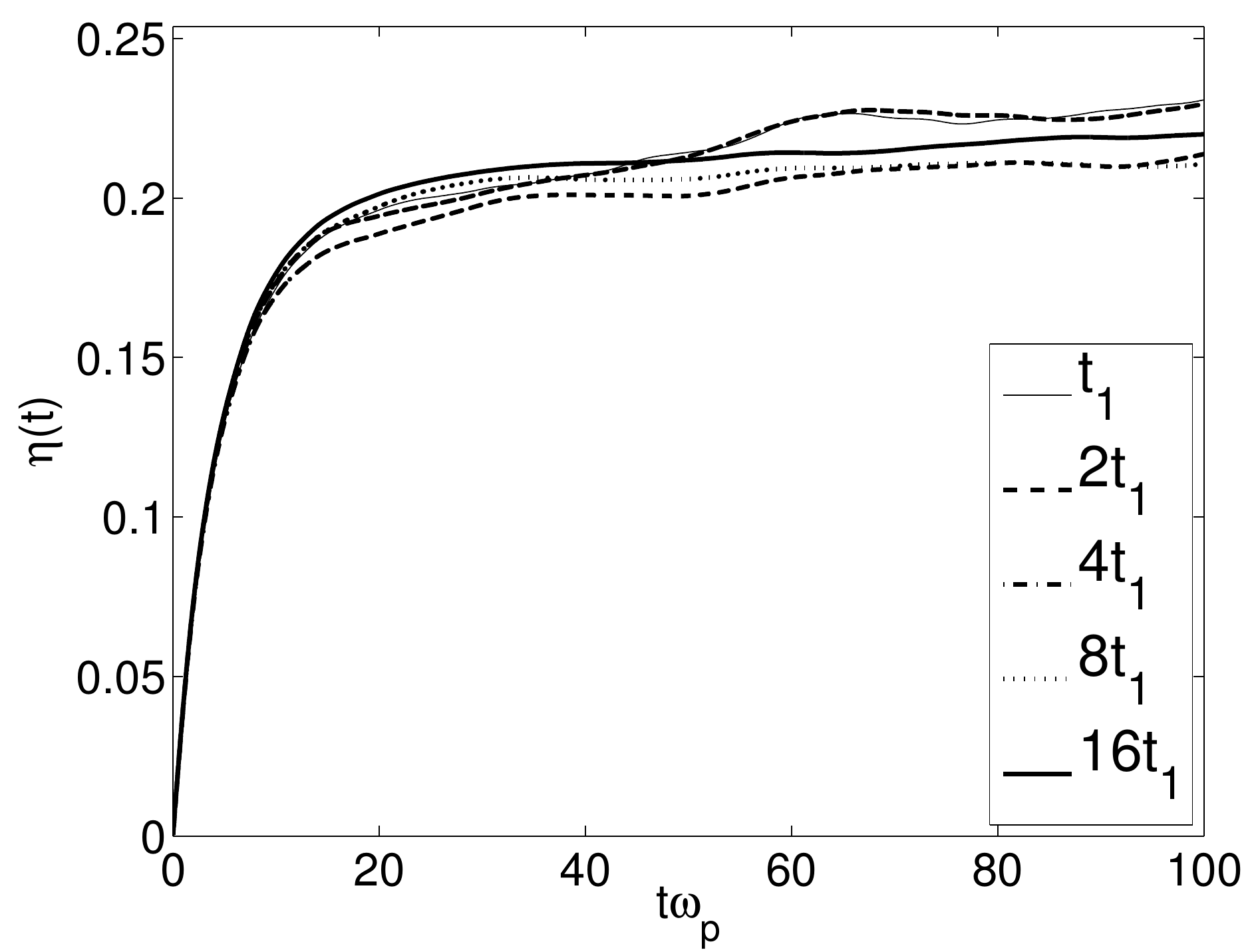}
\caption{Illustration of the calculation with equilibrium MD of the Green-Kubo expression (\ref{GK_viscosity}) for the shear viscosity coefficient.
Shown are results for the OCP with $\kappa=0$ and for three different values of the coupling parameter $\Gamma$: $\Gamma=0.5$ (upper row), $\Gamma=30$ (middle row), and $\Gamma=120$ (lower row). 
Results for the (dimensionless) shear-stress autocorrelation function $\sum_{x\neq y}^3\langle \sigma_{xy}(t)\sigma_{xy}(0)\rangle$ obtained using Eq.(\ref{numerical_correlation_function}) are shown in the left-hand column.
Corresponding (dimensionless) cumulated sums $\eta(t)$ defined by Eq.(\ref{cumulated_sum_eta_approximate}) are shown in the right-hand column.
In each figure, the results corresponding to five different simulation lengths are shown (in $1/\omega_p$ units); the other simulation parameters are listed in table \ref{table1}.
In the left-hand side figures, the curves have been shifted vertically for clarity (the horizontal dashed lines corresponds to the shifted zero of the vertical axis.)
}
\label{Fig1}
\end{figure*}
The simulations were performed as follows.
Initial particle positions were assigned randomly in the simulation box, with a small region surrounding each particle excluded to avoid initial explosion.
Initial particle velocities were assigned randomly from a Maxwell-Boltzmann distribution at the desired temperature.
The simulation time consisted of an equilibration phase of length $t_{eq}=N_{eq}\delta t$ followed by the main MD run of length $t_{run}=N_{run}\delta t$, for a total of $N_{eq}+N_{run}$ time steps.
During the equilibration phase, velocity scaling (also known as the Berendsen thermostat \cite{FrenkelSmit}) was used at every timestep to maintain the desired temperature.
Velocity scaling was turned off after the equilibration phase, at which point the simulations transitioned to the main MD run phase in which particle positions and velocities were recorded at every timestep.

The viscosity coefficient was evaluated from the Green-Kubo relation, Eq.~(\ref{GK_viscosity}), as follows.
First, $\sigma_{xy}(t)$ was computed using Eqs.~(\ref{sigma_kin})-(\ref{sigma_lr}) at each time step $t=n\delta t$, $0\leq n\leq N_{run}$ using the positions and velocities from the MD simulation.
Second, the shear stress autocorrelation function (\ref{shearstressautocorrelationfunction}) was computed by first replacing the thermal average by a time average
\be
J_{xy}(\tau)&=&\lim_{t\to\infty}{\bar{J}_{xy}(t,\tau)}\\
&\simeq& \bar{J}_{xy}(t_{run},\tau) \label{time_average}
\ee
where
\be
\bar{J}_{xy}(t,\tau):=\frac{1}{t-\tau}\int_0^{t-\tau}{\sigma_{xy}(s+\tau)\sigma_{xy}(s)ds} \label{time_average}
\ee
and then discretizing in time
\be
\lefteqn{\bar{J}_{xy}(t_{run}=N_{run}\delta t,\tau=n\delta t)}&&\label{numerical_correlation_function}\\
\!&\simeq&\!\frac{1}{N_{run}+1-n}\sum_{m=0}^{N_{run}-n}{\sigma_{xy}((m+n)\/\delta t)\sigma_{xy}(m\/ \delta t)}\nn .
\ee
Third, the cumulated sum
\be
\eta(\tau=n\delta t)&:=&\frac{1}{6Vk_B T}\sum_{x=1}^3{\sum_{y\neq x=1}^3{\int_0^\tau{\big\langle \sigma_{xy}(t)\sigma_{xy}(0) \big\rangle_{eq} dt}}}\nn\\
\label{cumulated_sum_eta}\\
&\simeq&\sum_{m=0}^{n}{ \frac{\delta t}{6Vk_BT}\sum_{x\neq y}^3{\bar{J}_{xy}(N_{run}\delta t,m\delta t)} } \label{cumulated_sum_eta_approximate}
\ee
was calculated. Ideally, according to the Green-Kubo relation (\ref{GK_viscosity}), the viscosity coefficient is given by $\eta=\lim_{t_{run}\to\infty} \eta(t_{run})$ (neglecting other systematic errors due to, e.g., size effects, $N$, force accuracy, etc.)
One expects the sum (\ref{cumulated_sum_eta}) to converge towards $\eta$ after a time longer than the correlation time scale of the correlation function.
Beyond that time, the correlation function vanishes and the cumulated sum reaches a plateau value equal to the viscosity coefficient.
In practice, as we shall see, the convergence to a plateau is quite slow and the accurate determination of the viscosity is impossible unless one performs very long simulations.

Figure~\ref{Fig1} illustrates the method and its convergence for a Coulomb OCP ($\kappa=0$) at three values of the coupling parameter across the fluid regime: $\Gamma=0.5, 30$ and $120$.
The figures on the left-hand side show the shear-stress autocorrelation function calculated using Eq.~(\ref{numerical_correlation_function}), those on the right-hand side show the cumulated sum Eq.~(\ref{cumulated_sum_eta_approximate}).
In each case, the results of simulations for four different simulation lengths are shown, namely $t_{run}=t_1,2\,t_1,4\,t_1,8\,t_1$ and $16\,t_1$ with $t_1=5242.88/\omega_p$, all the other numerical parameters being identical (see table~\ref{table1}).
The final results reported in Sec.~\ref{section_III} were all obtained
using $t_{run}=16 t_1$; note also that the shortest runs shown here
are in fact longer than the simulation duration $\sim
10^3-10^4/\omega_p$ previously reported in
the literature \cite{bast:05,SalinCaillol2003}.
For all $\Gamma$ values, the autocorrelation function decays toward zero on a time scale $t_c\sim{\rm 10's}/\omega_p$, which is much smaller than the simulation time scale since $t_c\ll t_1$.
Nevertheless, the length of the simulation has a significant effect on the noise and, in turn, on the convergence of the Green-Kubo calculation.
Thus, in all cases, the correlation function obtained with the shortest MD run $t_{run}=t_1$ never fully vanishes as time increases: it decays toward zero and then alternatively stays above and below the horizontal axis.
When integrated over time, this leads to a significant noise in the cumulated sum,  which never quite reaches a plateau value.

\subsection{Convergence Study}

In this section, we report on the analysis of the speed of convergence of the calculation that we have undertaken to select the numerical parameters of table~\ref{table1} used to calculate the viscosity coefficients reported in section~\ref{section_III}.
The goal is to empirically answer the question: how large should the simulation length $t_{run}$ be in order to attain the desired accuracy.

\subsubsection{Initial time behavior}

\begin{figure}[t!]
\includegraphics[width=\columnwidth]{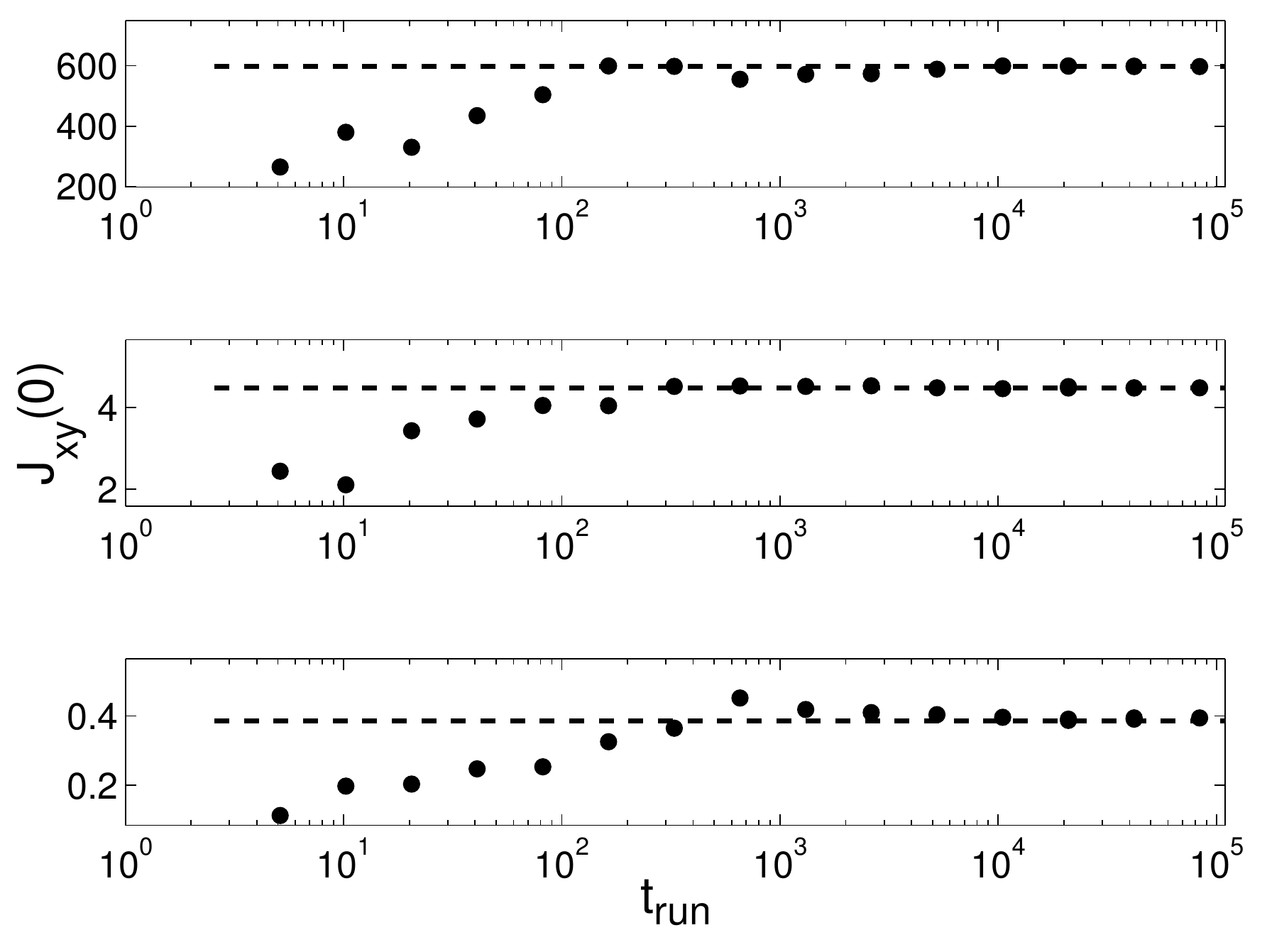}
\caption{
Illustration of the convergence of the sum rule (\ref{initialJxy}) with the length of the simulation $t_{run}$ for $\Gamma=1$ (upper panel), $\Gamma=20$ (middle panel), and 
$\Gamma=175$ (lower panel). In all cases $\kappa=0$.
Dashed lines represent Eq.~(\ref{initialJxy}) evaluated with the $g(r)$ obtained from the MD simulation, 
while the dots represents results of direct calculations of $J_{xy}(0)$ from the stress tensor.}
\label{Fig2}
\end{figure} 
We start with a calculation of the initial value of the correlation function: $J_{xy}(0)=\langle \sigma_{xy}(t)\sigma_{xy}(t)\rangle$.
Accuracy in the determination of $J_{xy}(0)$ is important since a shift in its value would certainly correspond to a shift of the entire time-evolution, which would lead to error in the cumulated sum and viscosity.
We find that although $J_{xy}(0)$ is less subject to statistical noise than $J_{xy}(t)$ for $t>0$, it is sufficient to cause concern.
Remarkably, a number of properties (or ``sum rules'') concerning $J_{xy}(0)$ are known and can be used to monitor the converge of its numerical determination.
In particular, the following exact sum rule can be shown
\be
J_{xy}(0)&=&N(k_BT)^2 \label{initialJxy}\\
&\!+\!&\!\!\frac{2\pi N nk_{B}T}{15}\!\int_0^\infty{\!\!\!\!\!dr r^3 \left(g(r)-\delta_{\kappa,0}\right)\left[4\phi^\prime(r)\!+r\!\phi^{\prime\prime}(r)\right]}\nn
\ee
Alternatively, using unitless quantities $\tilde{\sigma}=\sigma/(m(a\omega_p)^2)$ and $\tilde v(r)=v(r/a)/k_BT$,
\be
\tilde{J}_{xy}(0)&=&\frac{J_{xy}(0)}{[m(a\omega_p)^2]^2}\\
&=&\frac{N}{9\Gamma^2}\label{initial_tildeJxy}\\
&+&\frac{N}{90\Gamma}\int_0^\infty{dr r^3\left[g(r)-\delta_{\kappa,0}\right]\left[4\tilde{v}^\prime(r)+r\tilde{v}^{\prime\prime}(r)\right]} .\nn 
\ee

\begin{figure}[t!]
\includegraphics[width=\columnwidth]{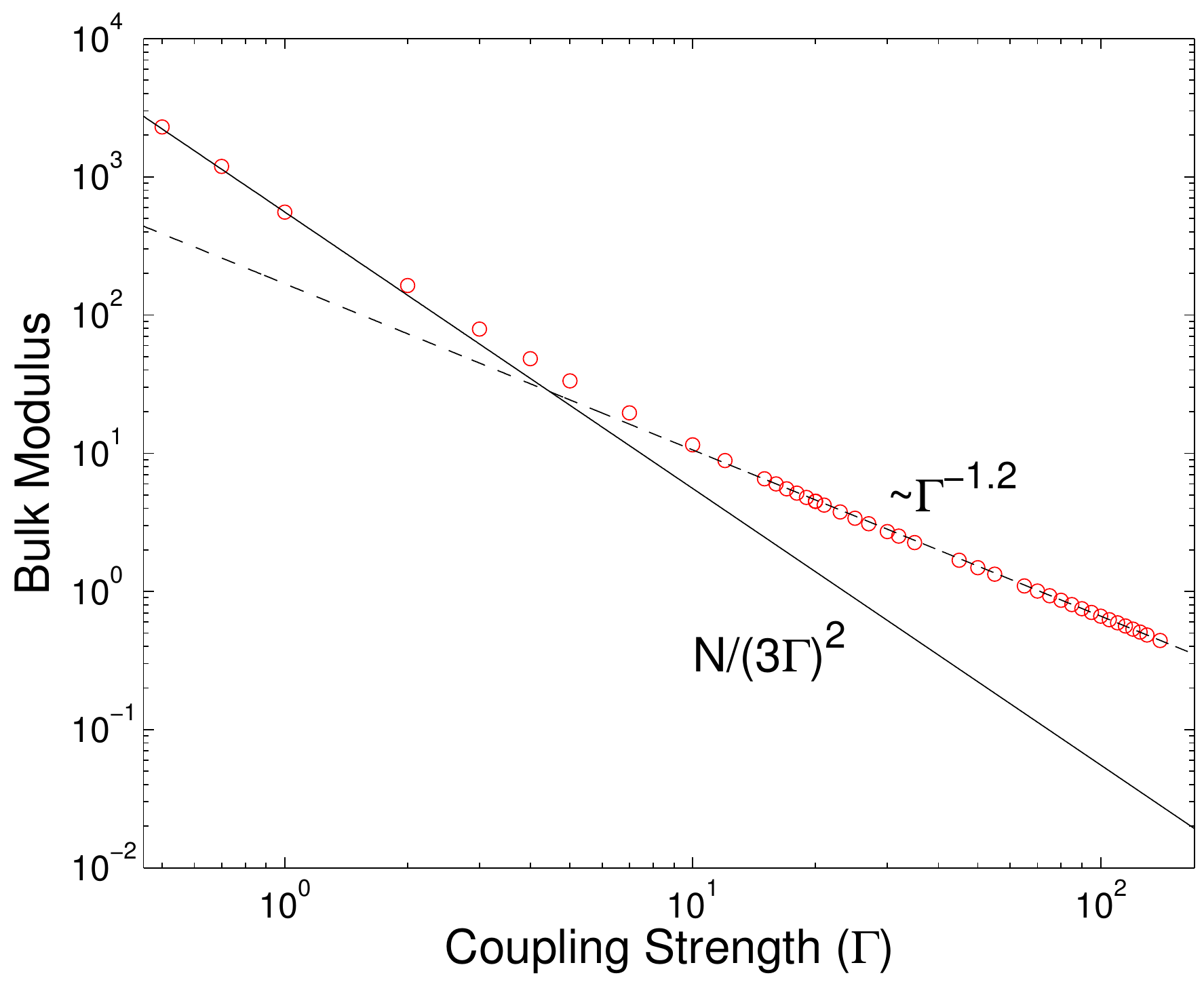}
\caption{
(color online) Bulk modulus of the Coulomb OCP as a function of $\Gamma$.
The vertical axis shows the dimensionless quantity $\tilde{J}_{xy}(0)=\frac{J_{xy}(0)}{[m(a\omega_p)^2]^2}=\frac{N}{(3\Gamma)^2}\frac{G}{nk_BT}$.}
\label{Fig2bis}
\end{figure} 
Figure~\ref{Fig2} shows a plot of the evolution of $J_{xy}(0)$ with the length of the simulation for a Coulomb OCP at $\Gamma=1,20$ and $175$.
In all cases, the initial value of the correlation function equals the expected value for simulation lengths greater than $t^*\sim 6000/\omega_p$.
The inaccuracy grows rapidly when the simulation length lies below this value.
The convergence of the correlation function at times $t\geq 0$ is discussed in the following subsection.

Figure~\ref{Fig2bis} shows the converged results for $J_{xy}(0)$ as a function of the coupling strength $\Gamma$.
Note that this quantity is simply related to the isothermal bulk modulus $G=-V\frac{dP}{dV}=n\frac{dP}{dn}$, such that
\be
J_{xy}(0)=V k_BT G .
\ee
In the weakly coupled regime, the equation of state is dominated by the ideal gas, kinetic pressure $P=nk_BT$, i.e. $G=nk_B T$ and $J_{xy}(0)=N(k_BT)^2$; accordingly $\tilde{J}_{xy}(0)\simeq \frac{N}{9\Gamma^2}$.
In the strongly coupled regime, the interaction energy dominates the pressure and the MD data show that  $\tilde{J}_{xy}(0)$ scales like $\sim \frac{N}{\Gamma^{1.2}}$.

\begin{figure}[t!]
\includegraphics[width=\columnwidth]{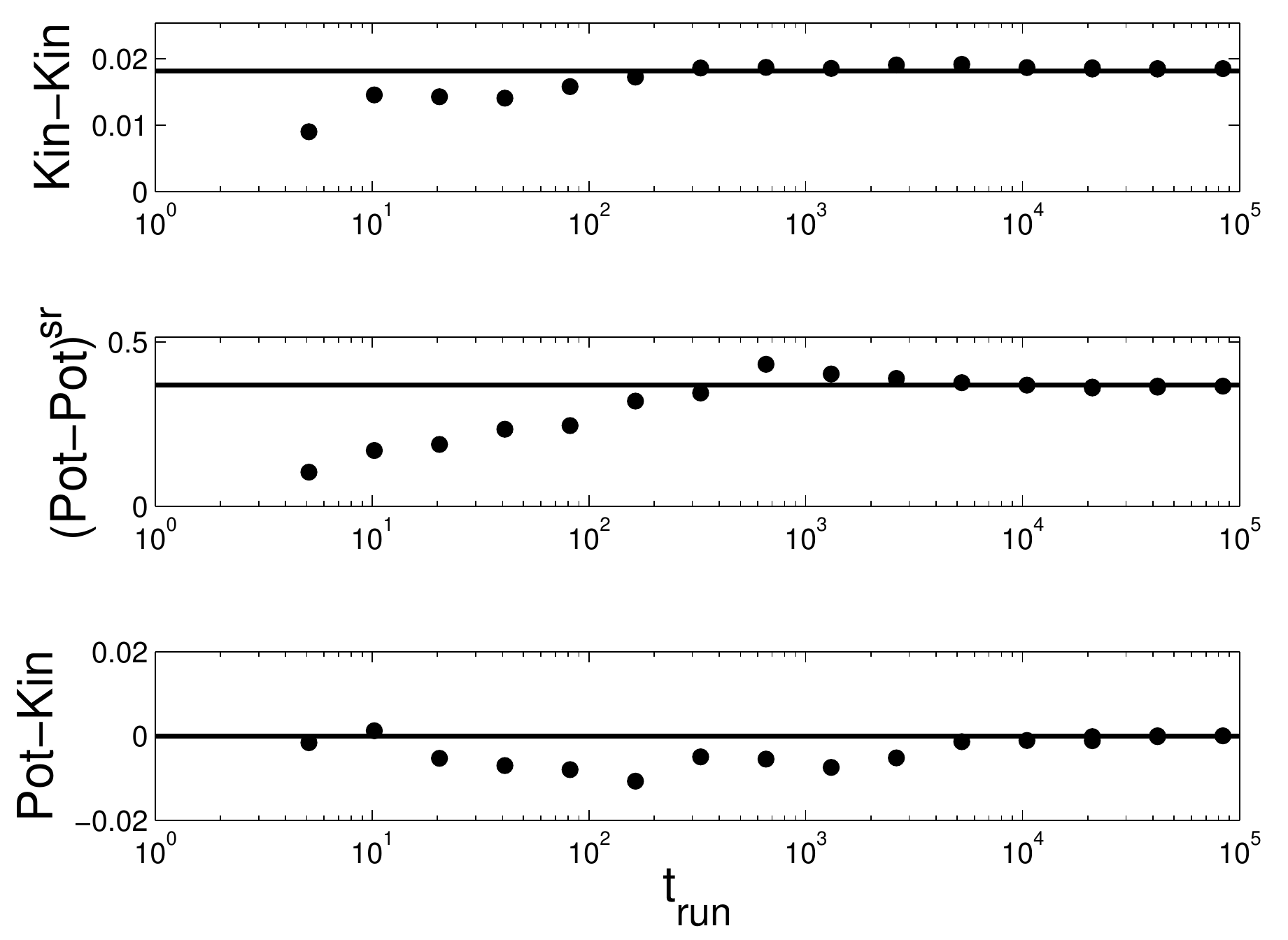}
\caption{
Illustration of the convergence of the detailed components of the sum rule (\ref{initialJxy}) with the length of the simulation $t_{run}$ for $\Gamma=2$ and $\kappa=0$.
Dashed lines represent the exact value, the dots represents results of direct calculations from the components of the stress tensor.
Top panel: kinetic-kinetic term (\ref{kin_kin}).
Middle panel: sr potential - potential term (\ref{srpot_pot}).
bottom panel: kinetic-potential term (\ref{kin_pot})
}
\label{Fig3}
\end{figure}

A more detailed study of the convergence of the initial value can be obtained by considering the different components of the sum rule (\ref{initialJxy}) obtained by substituting the decomposition (\ref{sigma_decomposition}) in Eq.(\ref{shearstressautocorrelationfunction}).
The first term in Eq.~(\ref{initialJxy}) corresponds to the kinetic-kinetic contribution
\be
J_{xy}^{kin}(0)=\big\langle \sigma_{xy}^{\rm kin}(0)\sigma_{xy}^{\rm kin}(0)\big\rangle_{eq}&=&N\,(k_BT)^2 . \label{kin_kin}
\ee
As shown in the appendix, the direct evaluation of $J_{xy}^{kin}(0)$ using Eq.~(\ref{numerical_correlation_function}) in an MD simulation amounts to calculating
\be
J_{xy}^{kin}(0)&\simeq&
\left[\frac{2}{t_{run}}\int_0^{t_{run}}{\sum_{i=1}^N{\frac{1}{2} mv_{x,i}(t)^2}dt}\right]\nn\\
&&\times\left[\frac{2}{t_{run}}\int_0^{t_{run}}{\sum_{i=1}^N{\frac{1}{2} mv_{y,i}(t)^2}dt}\right]\nn\\
&+&\text{cross term} . \label{Jxykin_tilde}
\ee
where the full expression for the cross term is given in Eq.(\ref{cross_term}).
In the large $t_{run}$ limit, the first term, which is related to the product of the averaged instantaneous kinetic energy $\sum_{i=1}^N{\frac{1}{2} mv_{x,i}(t)^2}$, is expected to converge to the exact result (\ref{kin_kin}), while the remaining cross term is expected to vanish.
In a MD simulation, the instantaneous kinetic energy fluctuates around the target velocity and the first term rapidly converges with $t_{run}$ to its limit value.
On the contrary, the cross term, which involves contributions that are quartic in the velocities, does not converge as fast as the kinetic energy to its limit value, which is quadratic in the velocities.
This is illustrated in Fig.~\ref{Fig3} (top panel): the value of (\ref{Jxykin_tilde}) converges to the expected value at large enough $t_{run}$ beyond which the cross terms are negligibly small compared to the first term.

We now discuss the term involving the interaction only,
\be
J_{xy}^{pot}(0)&:=&\big\langle \left[\sigma_{xy}^{\rm sr}(0)+\sigma_{xy}^{\rm lr}(0)\right]\left[\sigma_{xy}^{\rm sr}(0)+\sigma_{xy}^{\rm lr}(0)\right] \big\rangle_{eq}\label{pot_pot}\\
&=&\frac{2\pi N}{15}nk_{B}T\int_0^\infty{dr r^3 g(r)\left[4\phi^\prime(r)+r\phi^{\prime\prime}(r)\right]}\nn
\ee
The later can actually be further broken into two components; for instance \cite{noteDaligault},
\be
J_{xy}^{sr}(0)&:=&\big\langle \sigma_{xy}^{\rm sr}(0) \left[\sigma_{xy}^{\rm sr}(0)+\sigma_{xy}^{\rm lr}(0)\right] \big\rangle_{eq}\label{srpot_pot}\\
&=&\frac{2\pi N}{15}nk_{B}T\int_0^\infty{dr r^3 g(r)\left[4\phi_{\rm sr}^\prime(r)+r\phi_{\rm sr}^{\prime\prime}(r)\right]} . \nn\\ 
\label{srpot_pot_exact}
\ee
Figure~\ref{Fig3} shows the convergence of $J_{xy}^{sr}(0)$ with the simulation length $t_{run}$ toward the exact value Eq.(\ref{srpot_pot_exact}).
Again we find that long $t_{run}$ must be used to ensure convergence.

Finally, the kinetic-potential term
\be
J_{xy}^{kin-pot}(0)&:=&\big\langle \sigma_{xy}^{\rm kin}(0) \left[\sigma_{xy}^{\rm sr}(0)+\sigma_{xy}^{\rm lr}(0)\right] \big\rangle_{eq}\label{kin_pot}\\
&=&\big\langle \left[\sigma_{xy}^{\rm sr}(0)+\sigma_{xy}^{\rm lr}(0)\right]\sigma_{xy}^{\rm kin}(0) \big\rangle_{eq}\nn\\
&=&0\,.
\ee
In practice, the term is negligibly small for long enough simulation length $t_{run}$.

In conclusion, the initial value of the correlation function converges relatively slowly towards its expected value; simulations longer than $t^*$ are necessary to reproduce the expected value.
The slow convergence is found to be caused by cross terms that vanish in the ideal limit but are finite in practice.

\subsubsection{Finite time correlation function}

The initial time correlation function determines a lower bound, $t^*$, for the simulation length needed to calculate the viscosity coefficient.
However, Fig.~\ref{Fig1} shows that this is far too short to obtain the accurate correlation functions necessary to evaluate the viscosity coefficient.
The intermediate time dynamics of $J_{xy}(t)$ does not converge as fast as the short-time dynamics, which leads to large variations in the evaluation of the viscosity coefficient.
As a consequence, the cumulated sum does not reach a plateau value at time $t^*$.
It is noteworthy that the time $t^*$ is actually larger than the simulation lengths used in previous studies \cite{SalinCaillol2003}.
Figure~\ref{Fig1} reveals that satisfying convergence can be achieved for simulation times on the order of $\sim 16 t^*$, corresponding to over $8$ million time steps for $\delta t=0.01/\omega_p$.

In order to understand this behavior, we employ the statistical error analysis of Zwanzig and Ailawadi \cite{ZwanzigAilawadi1969,FrenkelSmit}.
Zwanzig and Ailawadi gave an error estimate for the deviation between the shear-stress autocorrelation function at time $t$ obtained with an MD simulation of finite length $t_{run}$, and its exact value $\bar{J}_{xy}(\infty,t)$ 
\be
\Delta(t)=\bar{J}_{xy}(t_{run},t)-\bar{J}_{xy}(\infty,t) ;
\ee
see Eq.~(\ref{time_average}).
To this end, they assumed that $\sigma_{xy}(t)$ is a Gaussian random variable (average denoted by $\langle \dots\rangle$ below), which was shown to give the correct order of magnitude of error estimates \cite{Bitsanis1987,FrenkelSmit}.
Under this assumption, they arrived at the result
\be
\langle\Delta(t_1)\Delta(t_2)\rangle&\simeq& \frac{2\tau_c}{t_{run}}\bar{J}_{xy}(\infty,0)^2\label{Zwanzig_result}\\
&&\text{with }0\leq t_1,t_2\leq \tau_c\ll t_{run}\nn
\ee
where
\be
\tau_c=2\frac{\int_0^\infty{dt\,J_{xy}(t)^2}}{J_{xy}(0)^2}
\ee
measures the relaxation time within which the exact correlation function $\bar{J}_{xy}(\infty,t)$ decays to zero from its initial value.
Applying Eq.~(\ref{Zwanzig_result}) with $t=t_1=t_2$ shows that the absolute error $\langle\Delta(t)^2\rangle$ in $J_{xy}(t)$ is independent of $t$. Therefore, the relative error
\be
\ds\frac{\left\langle\left[\bar{J}_{xy}(t_{run},t)-\bar{J}_{xy}(\infty,t)\right]^2\right\rangle}{\bar{J}_{xy}(\infty,t)^2}&\simeq& \frac{2\tau_c}{t_{run}}\left[\frac{\bar{J}_{xy}(\infty,0)}{\bar{J}_{xy}(\infty,t)}\right]^2\label{relative_error}
\ee
increases rapidly as $\bar{J}_{xy}(\infty,t)$ goes to $0$ \cite{FrenkelSmit}.
Equation (\ref{relative_error}) shows that this increase in the relative error can be lessened by increasing the simulation length $t_{run}$.
This is indeed consistent with the results in Fig.~\ref{Fig1} for $t_{run}=t_1$ and $t_{run}=2t_1$, although the initial value of the correlation function has converged to its expected value, the values at later times $t\leq \tau_c$ are not converged.
This noise is reduced when the simulation length is increased to $16t_1$.

\section{Final Results} \label{section_III}

\begin{figure}[t!]
\includegraphics[width=\columnwidth]{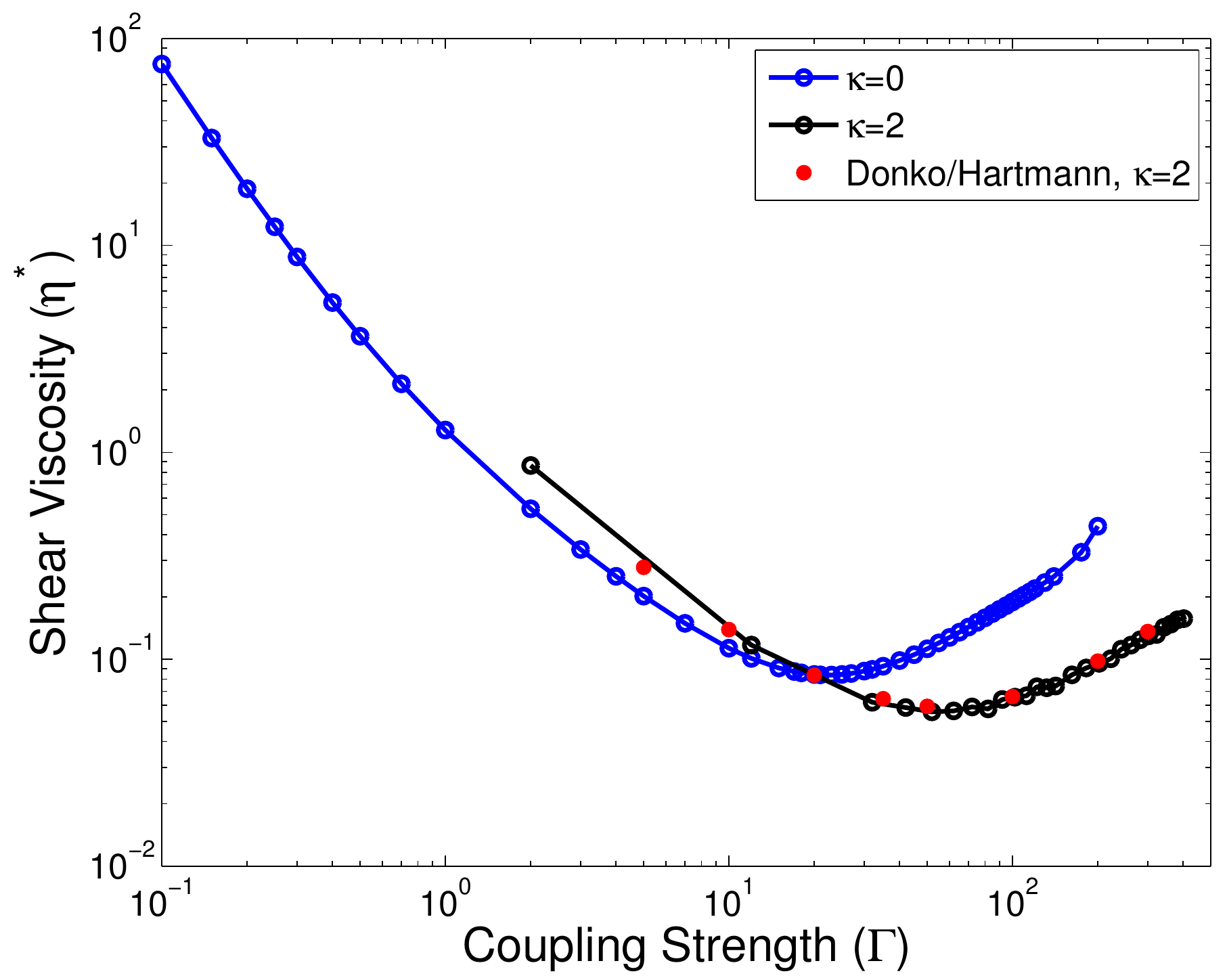}
\caption{(color online) Shear viscosity coefficient across the fluid phase of the one-component plasma at $\kappa=0$ (blue open dots) and at $\kappa=2$ (black full dots) obtained with the equilibrium MD simulations described in the text. The lines between the dots are included to guide the eyes.
At $\kappa=2$, the red dots show the non-equilibrium MD results of Donk{\'o} and Hartmann \cite{DonkoHartmann2008}.
}
\label{Fig4}
\end{figure}

\begin{table}[t]
\begin{tabular}{c|c||c|c||c|c}
$\Gamma$ & $\eta/(mna^2\omega_p)$ & $\Gamma$ & $\eta/(mna^2\omega_p)$ &
$\Gamma$ & $\eta/(mna^2\omega_p)$ \\\hline
    0.1   &    75.2  &    20  &  0.084     & 75  &  0.152 \\
    0.5  &  3.6907   &  21   & 0.084    & 80  &  0.160 \\
    0.7  &  2.1546   &  23 &   0.085  &   85  &  0.168 \\
    1   & 1.1831   & 25  &  0.085  &   90  &  0.176   \\
    2  &  0.4440  & 27 &   0.086  &   95 &   0.184   \\
    3   & 0.2755  &30   & 0.088   &  100  &  0.191  \\
    4   & 0.1928   & 32   & 0.089  &  105 &   0.199  \\
    5   & 0.1713  & 35  &  0.092 &  110  &  0.207  \\
   7   & 0.1345  &  40   & 0.097  & 115  &  0.214  \\
   10  &  0.101  & 45  &  0.102 &  120 &   0.222  \\
   12   & 0.0953  & 50 &   0.110 &  130  &  0.236  \\
   15  &  0.0864  &   55   & 0.119 &    140  &  0.251  \\
   17  &  0.083    &  60   & 0.128  & 175  &  0.3209\\
18  &  0.0830 &   65 &   0.136  & 200  &  0.4400 \\
19  &  0.0810 &    70 &   0.144  &  &
\end{tabular}
\caption{Shear viscosity coefficient $\eta$ of the one-component
  plasma with $\kappa=0$ at various coupling parameters $\Gamma$ as obtained with the
  molecular dynamics simulations described in the main text. Data are shown in units of $mna^2\omega_p$.}
\label{table2_kappa0}
\end{table}
The shear viscosity coefficients obtained using the method described in section \ref{section_II}, along with the numerical parameters collected in table~\ref{table1}, are shown in figure~\ref{Fig4} for $\kappa=0$ and $\kappa=2$. Also shown are the data of Donk{\'o} and Hartmann, obtained from non-equilibrium MD for $\kappa=2$ (see table I in \cite{DonkoHartmann2008}) and the data of Bastea obtained from equilibrium MD for $\kappa=0$ (we plot the fitting formula (11) of \cite{bast:05}).
For convenience, the numerical values are given in table~\ref{table2_kappa0} for $\kappa=0$ and in table~\ref{table2_kappa2} for $\kappa=2$.

We highlight the following important features of the present results.

\begin{figure}[t!]
\includegraphics[width=\columnwidth]{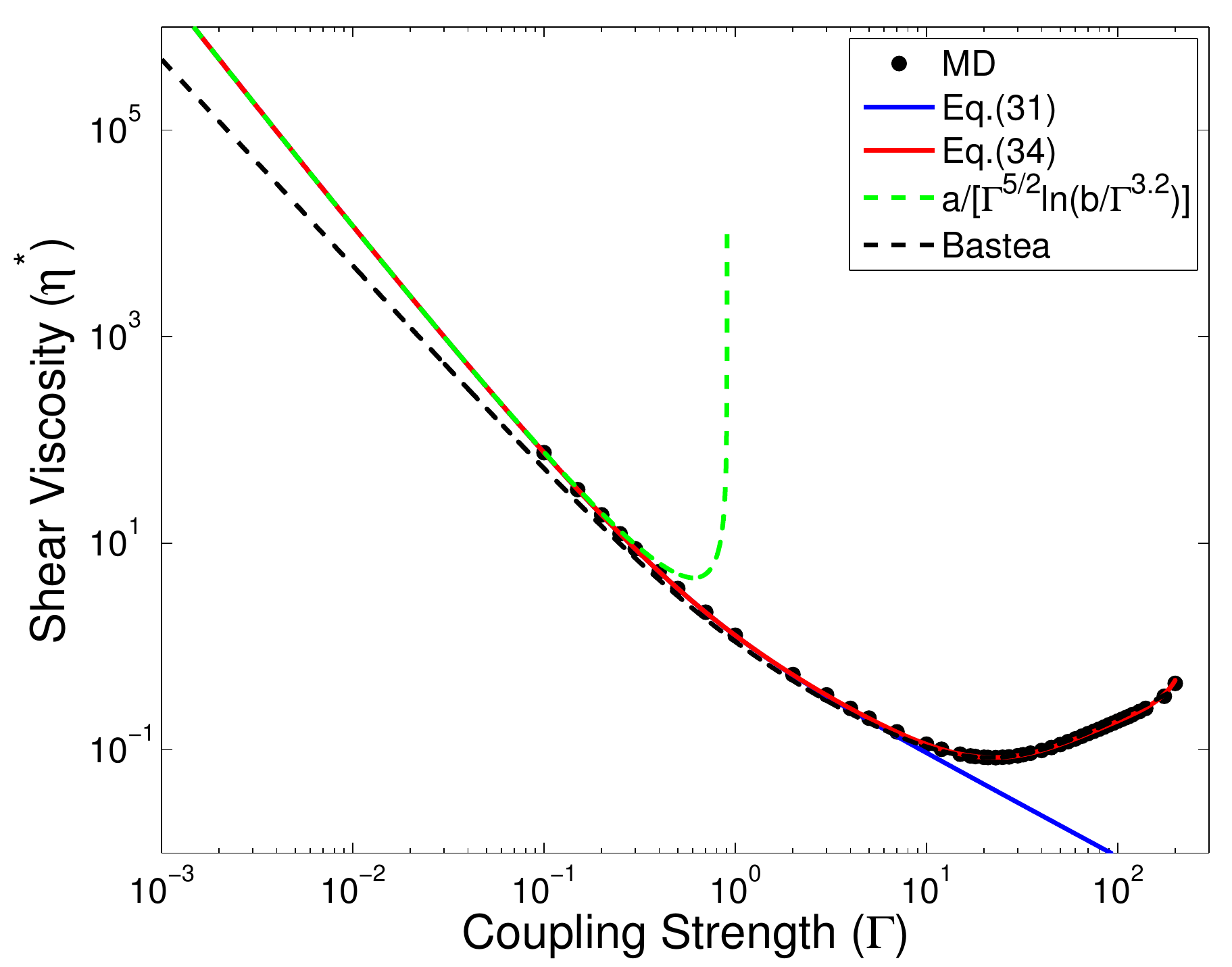}
\caption{(color online) Comparison of the MD data (dots) with the fitting formula Eq.(\ref{Practical_Formula}). The red line shows the full expression (\ref{Practical_Formula}), the green dashed line shows the LS limit $\eta^*(\Gamma)=\frac{a}{\Gamma^{5/2}\ln\left(\frac{b}{\Gamma^{3/2}}\right)}$, the blue line shows Eq.(\ref{extended_LS}), i.e. $\eta^*(\Gamma)=\frac{a}{\Gamma^{5/2}\ln\left(1+\frac{b}{\Gamma^{3/2}}\right)}$ , the black dashed line shows the formula of Bastea \cite{bast:05}.}
\label{Fig5}
\end{figure}

\noindent (1) In Fig.~\ref{Fig4}, the data at $\kappa=2$ are compared with the results of Donk{\'o} and Hartmann obtained using two independent non-equilibrium molecular dynamics calculations \cite{DonkoHartmann2008}.
We find very good agreement between all three independent calculations.
For $\kappa=0$, our data are in very good agreement with Bastea's fit, which was obtained by interpolating MD data over $0.05\leq\Gamma\leq 100$  \cite{bast:05}; see Fig.~\ref{Fig5}. 

\noindent (2) As shown in Fig.~\ref{Fig5}, for all $\Gamma<10$, the viscosity coefficient of the coulomb OCP ($\kappa=0$) is well approximated by
\be
\eta=\eta_0\frac{\delta}{\ln\left(1+C\frac{\lambda_D}{r_L}\right)}\label{extended_LS}
\ee
where $\ds \eta_0=\frac{5}{4}\sqrt{\frac{m}{\pi}}\frac{(k_B T)^{5/2}}{q^4}$, $\lambda_D=\sqrt{4\pi q^2 n/k_BT}$ is the Debye length, and $r_L=q^2/k_BT$ is the so-called distance of closest approach.
Here $\delta=0.466$ and $C=1.493$ are numerical parameters determined by interpolating the numerical data.
The model (\ref{extended_LS}) represents a straightforward modification of the traditional Landau-Spitzer (LS) formula \cite{Spitzerbook}
\be
\eta_{LS}=\eta_0\frac{1}{\ln \left(\frac{\lambda_D}{r_L}\right)} \label{eta_LS}
\ee
derived for weakly-coupled plasmas.
Indeed, in the weakly-coupled limit, Eq.(\ref{extended_LS}) reduces to $\delta\eta_0/\ln\left(C\frac{\lambda_D}{r_L}\right)$.
In the LS theory, the Coulomb logarithm $\ln \left(\frac{\lambda_D}{r_L}\right)$ arises from the long-range nature of the Coulomb force. It is usually expressed in terms of the Debye length $\lambda_D$ (which represents the largest impact parameter beyond which interactions are screened out), and of the distance $r_L$ (which characterizes the smallest impact parameter).
Our MD simulations reveal that, while the LS theory provides the right scaling at $\Gamma<<1$, the model must be corrected through the coefficients $C$ and $\delta$ to match the data.
The coefficient $C$ is a correction to the somewhat arbitrary
parameters $\lambda_D$ and $R_L$, which can be predicted by more
advanced theories (\cite{BaalrudDaligault2013prl} and literature therein).
The prefactor $\delta$ is a correction to the fact that LS corresponds to a single Sonine polynomial approximation in the Chapman-Enskog solution of the plasma kinetic equation.
Figure~\ref{Fig5} shows that the modified LS result $\delta\eta_0/\ln\left(C\frac{\lambda_D}{r_L}\right)$ breaks down at $\Gamma\sim 0.1$, while the simple modification (\ref{extended_LS}) extends its validity to the moderately coupled regime up to $\Gamma\sim 10$.
Remarkably, the same extension of the LS theory was found to work as well for other transport processes, including the electron-ion temperature relaxation rate \cite{DimonteDaligault2008} and the diffusion coefficients in mixtures \cite{Daligault2012}.

\begin{table}[t]
\begin{tabular}{c|c||c|c||c|c}
$\Gamma$ & $\eta/(mna^2\omega_p)$ & $\Gamma$ & $\eta/(mna^2\omega_p)$ &
$\Gamma$ & $\eta/(mna^2\omega_p)$ \\\hline
2 &   0.8638  &  102 &   0.0654   &  242 &   0.1117\\
12 &   0.1170  &  112 &   0.0665    &  262 &  0.1170\\
32 &   0.0619   &  122 &   0.0736   &  282 &   0.1242\\
42 &   0.0584   &  132 &   0.0728    &  302 &   0.1296\\
52 &   0.05572   &  142 &   0.0742   &  322 &   0.1316\\
62 &   0.0562   &  162 &   0.0840   &  342 &   0.1426\\
72 &   0.05882   &  182 &   0.0906   &  362 &  0.1478\\
82 &   0.0575   &  202 &   0.0955   &  382 &   0.1550\\
92 &   0.0637   &  222 &   0.1003   &  402 &   0.1571
\end{tabular}
\caption{Shear viscosity coefficient $\eta$ of the one-component
  plasma with $\kappa=2$ at various coupling parameters $\Gamma$ as obtained with the
  molecular dynamics simulations described in the main text. Data are shown in units of $mna^2\omega_p$.}
\label{table2_kappa2}
\end{table}
\noindent (3) The curve $\eta^*(\Gamma)$ presents a shallow minimum that is located in the range $18\leq \Gamma\leq 20$.
A more precise determination of the minimum is not possible with the accuracy of the present data.

\noindent (4) At high $\Gamma$, the viscosity $\eta$ and the
self-diffusion coefficient $D$ satisfy the Stokes-Einstein relation
\be
\frac{\pi a}{k_B T}D\eta=0.087\,\pm 2\%\quad\text{for }\Gamma\geq 50\,,
\ee
for $\kappa=0$.
For a detailed discussion on the Stokes-Einstein and its physical interpretation, see \cite{Daligault2006}.

\noindent (5) Finally, we provide a practical fit that reproduces the
viscosity coefficient across coupling regimes, from the weakly coupled
regime to the solid-liquid transition, in the form
\be
\eta^*(\Gamma)&=&\frac{\eta}{mna^2\omega_p}\nn\\
&=&\frac{a}{\Gamma^{5/2}\ln\left(1+\frac{b}{\Gamma^{3/2}}\right)}
\frac{1+a_1\Gamma+a_2\Gamma^2+a_3\Gamma^3}{1+b_1\Gamma+b_2\Gamma^2+b_3\Gamma^3+ b_4\Gamma^4} \nn\\
\label{Practical_Formula}
\ee
In Eq.(\ref{Practical_Formula}),  we enforce the model (\ref{extended_LS}) valid for $\Gamma<10$ and approximate the remainder with a Pad{\'e} (rational fraction) approximation.
As seen in Fig.~\ref{Fig5}, the formula (\ref{Practical_Formula}) together with the parameters listed in table~\ref{table3} is very accurate across the entire fluid regime.

We also compare our fit (\ref{Practical_Formula}) with that proposed by Bastea \cite{bast:05}, namely $\eta^*=0.482/\Gamma^2+0.629/\Gamma^{0.878}+0.00188\Gamma$, and obtained by fitting his MD data over the range $0.05\leq\Gamma\leq 100$.
While the later is quite accurate at moderate and strong coupling, it fails to reproduce the traditional Landau-Spitzer behavior in the weakly coupled regime.

\begin{table}[b!]
\begin{tabular}{c|c|c|c|c}
$a$ & $b$ & $a_1$ & $a_2$ & $a_3$ \\\hline
0.794811 & 0.862151 & 0.0425698  & 0.00205782  & 7.03658e-05
\end{tabular}
\begin{tabular}{c|c|c|c}
$b_1$ & $b_2$ & $b_3$ & $b_4$ \\\hline
0.0429942 &  -0.000270798  & 3.25441e-06 & -1.15019e-08
\end{tabular}
\caption{Fitting parameters to be used in Eq.(\ref{Practical_Formula}).}
\label{table3}
\end{table}

\section{Comparison to theoretical models} \label{section_IV}

In the previous section, we compared the MD results with the seminal theory of Landau-Spitzer.
In this section, we test the validity of theories that have been developed to predict the viscosity coefficients of the Coulomb OCP ($\kappa=0$) in the moderately and strongly coupled regime, namely the theory of Vieillefosse-Hansen, the kinetic theories of Wallenborn-Baus and of Tanaka-Ichimaru, and the recent effective potential theory of Baalrud-Daligault.

\begin{figure}[b]
\includegraphics[width=8.0cm]{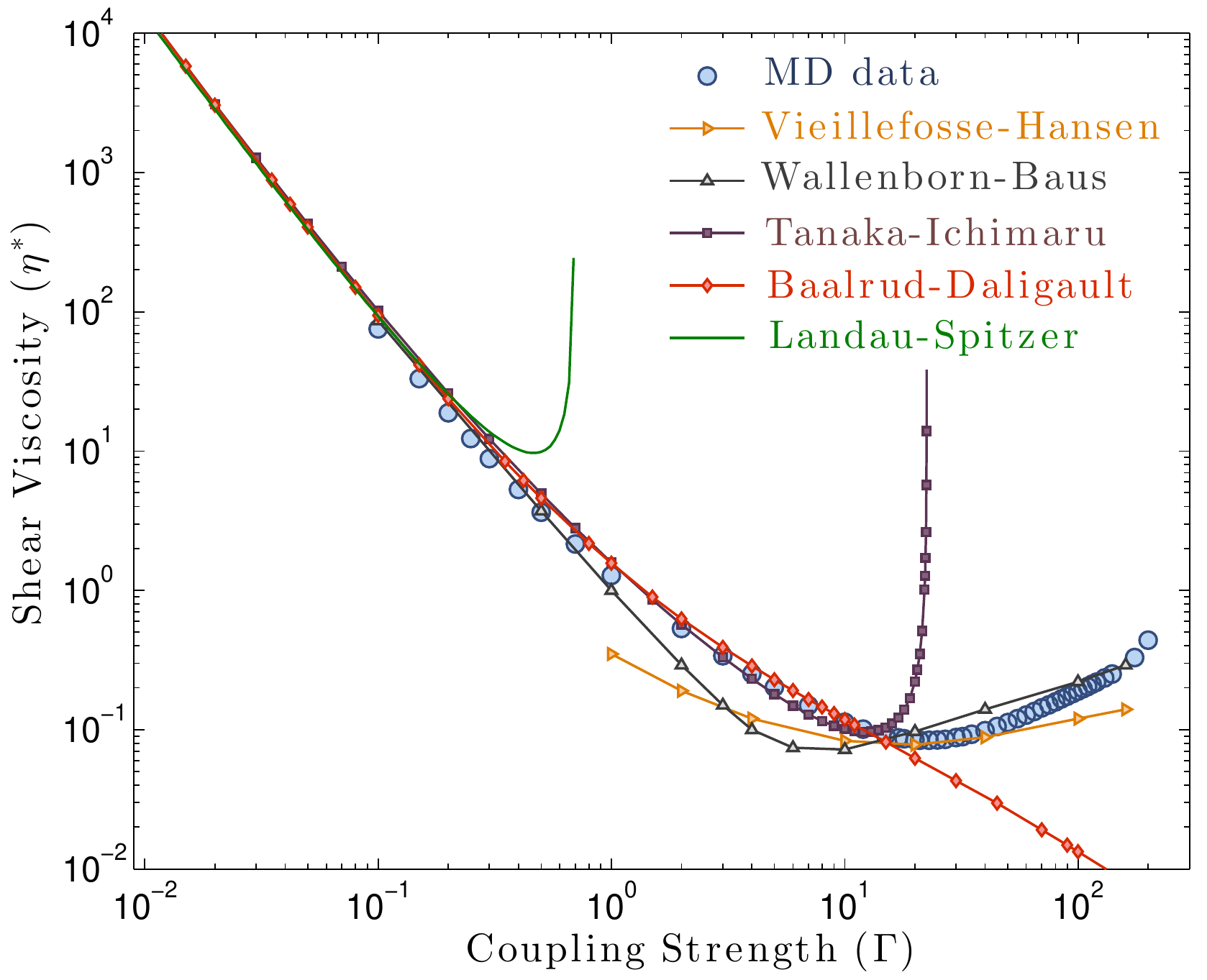}
\caption{(color online) Comparison of the MD results for the shear viscosity coefficient with the corrected Landau-Spitzer prediction discussed in Sec.\ref{section_III}, the theory of Vieillefosse-Hansen, the kinetic theories of Wallenborn-Baus and of Tanaka-Ichimaru, and the effective potential theory of Baalrud-Daligault.
See main text for a detailed comparison.
}
\label{Fig6}
\end{figure}

The predictions of these theories are compared with our new MD results in Fig.~\ref{Fig6}.
In the following, we briefly recall some basic facts about the various theories and discuss their validity with regard to the comparison with the MD results. 

\subsection{The Vieillefosse-Hansen theory}

Vieillefosse and Hansen \cite{VieillefosseHansen1975} applied the framework of the generalized hydrodynamics formalism.
Briefly, the known short-time expansion of the transverse-current autocorrelation function $C_{\perp}(k,t)$ up to fourth-order in time $t$ was used to build a Gaussian approximation of the memory function associated to $C_{\perp}$.
The coefficients of the Gaussian approximation depended on the first three frequency sum-rules of $C_{\perp}$ that can be exactly written in terms of the pair distribution function $g(r)$ and of the ternary distribution function $g_3(r,r^\prime)$.
Using the superposition approximation to express $g_3$ in term of $g$, the theory of Vieillefosse-Hansen depends on the pair distribution $g$ only.
Figure~\ref{Fig6} displays the results reported in Table III of the original paper \cite{VieillefosseHansen1975}.
Remarkably the predicted viscosity exhibits a minimum as a function of $\Gamma$ around $\Gamma=20$.
At $\Gamma=20$, Vieillefosse and Hansen give for the reduced viscosity coefficient $\eta^*=0.0781\pm0.004$, which is in good agreement with our MD result $\eta^*=0.084$ reported in Table~\ref{table2_kappa0}.
However, this good agreement may be fortuitous since, as seen in Fig.~\ref{Fig6}, the Vieillefosse-Hansen model greatly underestimate the viscosity at all other values of $\Gamma$.

\subsection{The Wallenborn-Baus theory}

Wallenborn and Baus applied the framework of renormalized equilibrium kinetic theory, a general kinetic theory of phase-space correlation functions, to derive an analytical model for the shear-viscosity coefficient \cite{wall:78}.
In this framework, the shear-viscosity coefficient can be exactly expressed in terms of the only unknown of the theory, the so-called generalized memory function.
They derived a sophisticated approximation for the latter that, by construction, attempts to account for (i.e. renormalize) the correlated motion of ions.
Their approximation reduces to the Lenard-Balescu collision operator when all the quantities involved in the memory function (e.g., the direct correlation function) are approximated by their weakly-coupled limiting values.
They then used their approximate memory function to calculate the shear-viscosity coefficient across coupling regimes.
The values of the shear-viscosity coefficient given in the original paper \cite{wall:78} are displayed in Fig.~\ref{Fig6}.
At weak coupling, the Wallenborn-Baus theory agrees with the MD data, which is consistent with the fact that the theory reduces to the Lenard-Balescu result with corrections due to short-range correlations, which determine the correction factor $C$ in the Coulomb logarithm (see Sec.~\ref{section_III}).
This theory does predict a minimum of the reduced viscosity coefficient with a value $\eta^*=0.007$ in fair agreement with the simulations, but at a coupling strength  $\Gamma\approx 8$, which is below the MD value of $\simeq 20$.

\subsection{The Tanaka-Ichimaru theory}

Tanaka and Ichimaru obtained a model for the shear viscosity coefficient by applying the framework of non-equilibrium kinetic theory, i.e. a theory for the temporal evolution of the non-equilibrium single-particle phase-space distribution functions $f({\bf r},{\bf p},t)$.
Using quasi-linear theory, they postulate an expression for the collision operator by introducing the notion of static local field correction $G(k)$, a quantity that accounts for static correlations between particles.
Their collision operator is
\be
\lefteqn{C_I(f,f) = \pi m \int \frac{d^3k}{(2\pi)^3} {\bf k} \cdot \frac{\partial}{\partial {\bf p}} \int d^3p^\prime \frac{v^2(k) [1-G(k)]}{| \epsilon ({\bf k} , {\bf k} \cdot {\bf p}/m)|^2}}&&\nn\\
&\times&\delta [{\bf k} \cdot ({\bf p} - {\bf p}^\prime)] {\bf k} \cdot \biggl( f ({\bf p}^\prime) \frac{\partial f}{\partial {\bf p}} - f({\bf p}^\prime) \frac{\partial f}{\partial {\bf p}^\prime} \biggr)\,, \label{C_TI}
\ee
where $\epsilon({\bf k}, \omega) = 1 - v(k) [1 - G(k)] \chi^{(0)} ({\bf k}, \omega)$ is the plasma dielectric function, $v(k) = \frac{4\pi e^2}{k^2}$, and $\chi^{(0)} ({\bf k}, \omega) = - \int d^3p \frac{{\bf k} \cdot \partial F/\partial {\bf p}}{\omega - {\bf k} \cdot {\bf v}}$ the density response function of the ideal gas, and $F$ the Boltzmann distribution function at temperature $T$ and density $n$.
In traditional weakly coupled plasma physics, correlations are neglected, i.e. $G(k)$ is set to zero, and Eq.~(\ref{C_TI}) reduces to the Lenard-Balescu collision operator.
By applying the Chapman-Enskog method to lowest order in the Sonine polynomial expansion, the following expression for the viscosity coefficient can be obtained \cite{IchimaruVol1}
\be
\eta_{TI} &=& \eta_0\frac{1}{\Xi_{TI}} .
\ee
Here, the generalized Coulomb logarithm 
\be
\Xi_{TI}= \frac{2}{\sqrt{\pi}} \int_0^\infty dk \frac{[1-G(k)]}{k} \int_0^\infty dz \frac{e^{-z^2}}{|\epsilon(k, k v_Tz)|^2}  \label{generalized_coulomb_logarithm}
\ee
arises, where $v_T = \sqrt{k_BT/m}$. This can be compared with Eq.~(\ref{eta_LS}).

Tanaka and Ichimaru have presented results for $\eta_{TI}$ using Eq.~(\ref{generalized_coulomb_logarithm}) with a local-field correction obtained by solving the hypernetted chain (HNC) equations with the bridge function correction of Ichimaru \cite{tana:86,IchimaruVol1}. The HNC equation gives access to the direct correlation function $c(k)$, which provides $G(k)= 1 + \frac{k_B T}{v(k)} c(k)$.
Reference \cite{IchimaruVol1} provides results for $0.1\leq \Gamma \leq 20$.
We have evaluated $\eta_{TI}$ using the same HNC equations, including Ichimaru's bridge function, for a wider range of values; see Fig.~\ref{Fig6}. This method agrees well with the MD data for $\Gamma \lesssim 10$. At $\Gamma \simeq 22.4$, $\Xi_{TI}$ crosses from positive to negative values, leading to a divergence of $\eta_{TI}$. 

The Tanaka-Ichimaru theory reduces to traditional plasma physics results in the weakly coupled limit. The simplest (Landau-Spitzer) plasma limit can be obtained by setting $G(k)\equiv 0$ (i.e., no correlations) and $\epsilon(k,\omega) \equiv 1$ (i.e., no screening). Then, $\Xi_{TI}$ reduces to the traditional Coulomb logarithm $\ln\Lambda=\ln \left(\frac{\lambda_D}{r_L}\right)$ when the usual cutoffs $\lambda_D$ and $r_L$ (see Sec.~\ref{section_III}) are used to regularize the $k$-integral in Eq.~(\ref{generalized_coulomb_logarithm}).
The Lenard-Balescu result is obtained by setting $G(k)\equiv 0$ but keeping the dielectric function $\tilde{\epsilon}$.
In this case, the $k$-integral converges at $k=0$ but a cutoff is necessary to regularize the remaining divergence at $k=\infty$.
This case was worked out by Braun \cite{Braun1967}, who expressed the result as a correction to the Landau-Spitzer viscosity coefficient as
\be
\eta_{LB}=\frac{\eta_{LS}}{1+ 0.346/\ln \left(\frac{\lambda_D}{r_L}\right)}.
\ee

\begin{figure}
\begin{center}
\includegraphics[width=7.5cm]{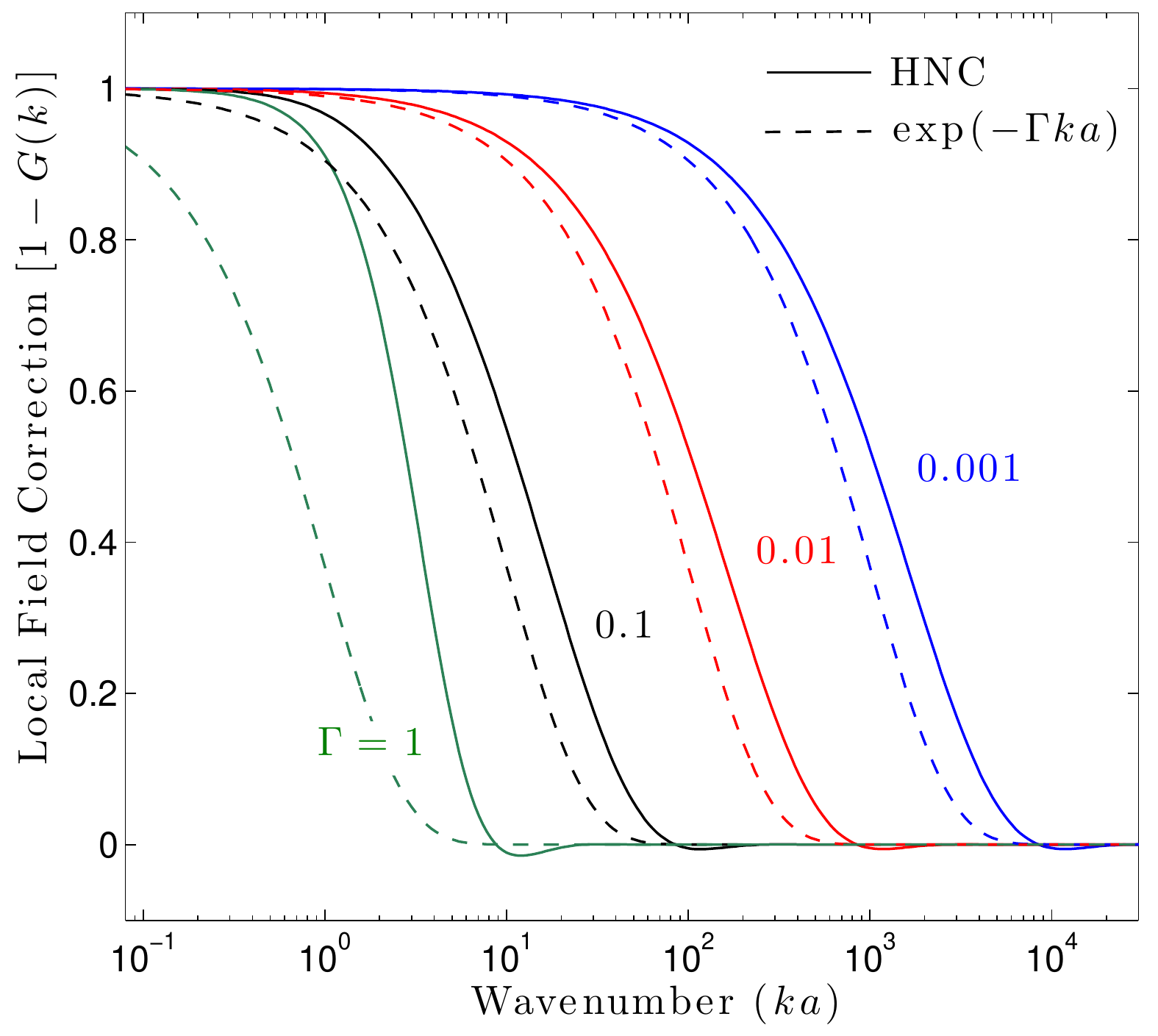}
\caption{(color online) Comparison between the local field correction obtained from the hypernetted chain (HNC) approximation, and Eq.~(\ref{eq:wclfc}) for weakly coupled OCP.}
\label{fg:lfc_wc}
\end{center}
\end{figure}

Alternative to these cutoffs, simple results for the local field correction can be obtained in the weakly coupled limit that allow analytic evaluation of the convergent integral in Eq.~(\ref{generalized_coulomb_logarithm}). Figure \ref{fg:lfc_wc} shows that 
\begin{equation}
1 - G(k) = \exp(-\Gamma ka)  \label{eq:wclfc}
\end{equation}
provides a good approximation for the OCP local field correction in the weakly coupled limit. If we also take the static dielectric function $\tilde{\varepsilon} = 1 + 3\Gamma [1-G(k)]/(ka)^2$ and note that the local field correction is negligible in this for weakly coupled plasmas [$\tilde{\varepsilon} \simeq 1 + 3\Gamma/(ka)^2$], we find
\begin{align}
\Xi_{TI} & \simeq \int_0^\infty d \bar{k} \frac{\bar{k} \exp(-\Gamma \bar{k})}{\bar{k}^2 + 3 \Gamma}  \label{eq:xiti_wc} \\ \nonumber
&= \frac{1}{2} \biggl[ E_1 (i/\Lambda) e^{i/\Lambda} + E_1 (-i/\Lambda) e^{-i/ \Lambda} \biggr]
\end{align}
in which $\Lambda = 1/(\sqrt{3} \Gamma^{3/2}) = \lambda_D/r_L$ is the OCP plasma parameter and $E_1$ is the exponential integral. Expanding for $\Lambda \gg 1$ gives
\begin{equation}
\Xi_{TI} \rightarrow \ln \Lambda - \gamma + \mathcal{O}{(\Lambda^{-1})} \label{eq:xiti_lim}
\end{equation}
where $\gamma$ is Euler's constant.  Note that Eq.~(\ref{eq:xiti_lim})
is the same result, including the order unity correction, as has been
obtained from other methods, including using the screened Coulomb
potential in the effective potential theory
(\cite{BaalrudDaligault2013prl} and references therein).


\subsection{The effective potential theory}

Recently, we proposed another approach for extending traditional plasma transport theories into the strong coupling regime \cite{BaalrudDaligault2013prl,BaalrudDaligault2014pop}.
Like traditional plasma theories, this is based on a binary scattering approximation, but where physics associated with many body correlations is included through the use of an effective interaction potential. This effective interaction potential was related to the potential of mean force, which is the interaction potential between two particles taking all surrounding particles to be at fixed positions. Like the other theories previously discussed, this also requires only the pair-distribution as input. Figure~\ref{Fig6} shows that this approach is accurate across coupling regimes up to approximately the minimum in the viscosity coefficient. 

\begin{figure}
\begin{center}
\includegraphics[width=8.2cm]{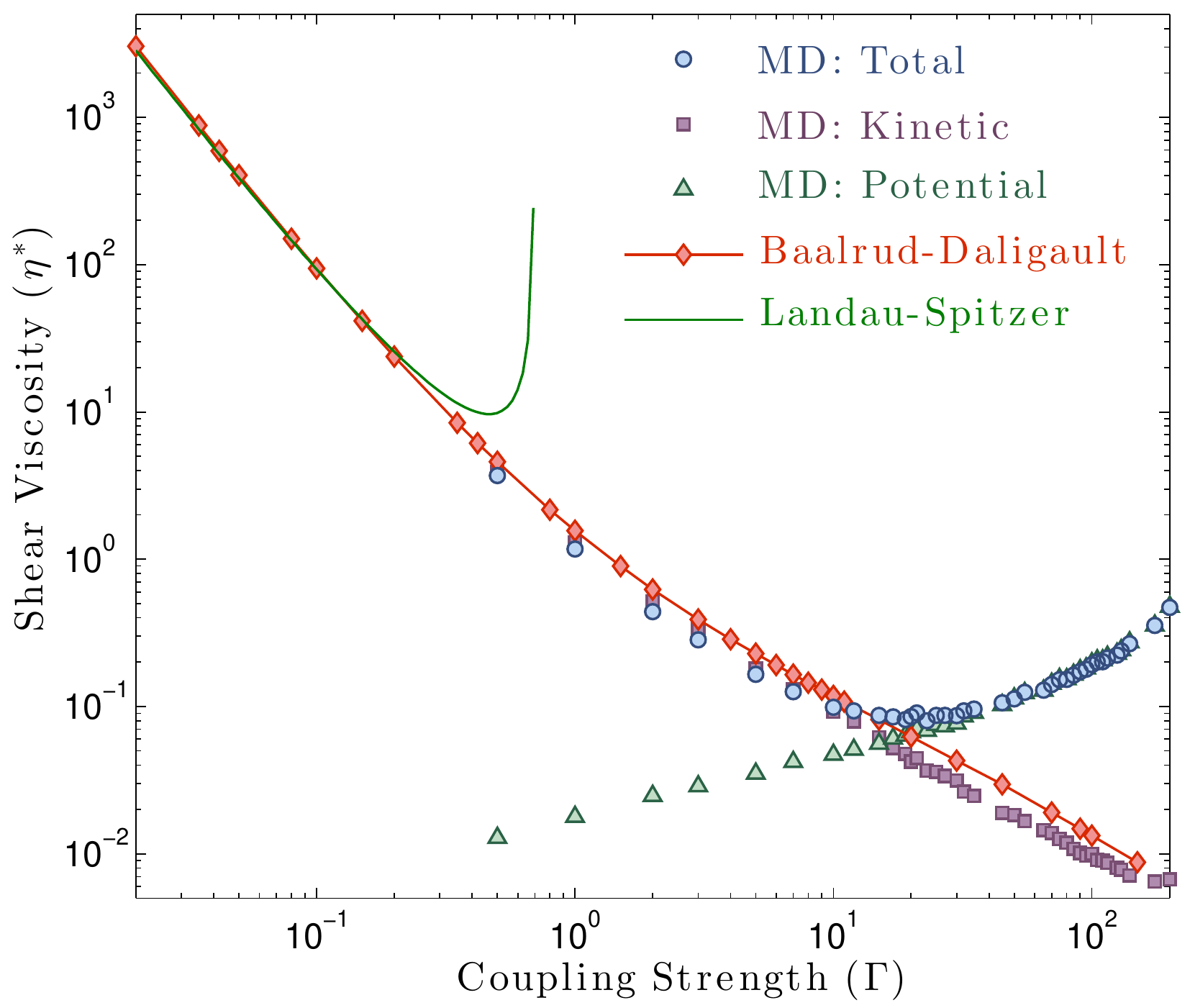}
\caption{Contributions to the viscosity coefficient computed from MD for $\kappa=0$ (for finite $\kappa$ values, see \ref{BausHansen1980}): kinetic (squares), potential (triangles), and total (circles). Also shown is the prediction of the effective potential theory (diamonds) and Landau-Spitzer theory (green line). }
\label{fg:eta_ichi}
\end{center}
\end{figure}

Breakdown of the effective potential theory arises at sufficiently strong coupling that the potential component of the viscosity dominates. This is expected because transport theories based on binary collisions only account for changes in the particle momenta, so they can at most describe the kinetic contribution.  This is shown in detail in Fig.~\ref{fg:eta_ichi}. 
This figure shows the kinetic-kinetic and potential-potential terms of the viscosity computed from MD using components of $J_{xy}(t)$ based on $\tensor\sigma^{\,{\rm kin}}$ and $\tensor\sigma^{\,{\rm pot}} = \tensor\sigma^{\,{\rm sr}} + \tensor\sigma^{\,{\rm lr}}$. 
We found that the cross terms (kinetic-potential and potential-kinetic) were negligible across the domain. 

For the theoretical evaluation, the viscosity was computed from the Chapman-Enskog relation
\begin{equation}
\eta^*_1 = \frac{5\sqrt{\pi}}{3\sqrt{3} \Gamma^{5/2} \Xi^{(2,2)}}
\end{equation}
where $\Xi^{(2,2)}$ was obtained using the method of \cite{BaalrudDaligault2013prl,BaalrudDaligault2014pop} inputing a pair distribution function calculated from the HNC approximation (no bridge function was included for the HNC computations used here). Figure~\ref{fg:eta_ichi} shows that this theory accurately tracks the kinetic-kinetic term, but contains no information about the potential-potential term. This is similar to how binary collision operators predict only the ideal gas component of the equation of state, whereas an additional term dependent on the pair distribution is required to describe the potential contribution at strong coupling. The effective potential theory breaks down at sufficiently strong coupling even for transport coefficients that do not have potential components, such as diffusion or temperature relaxation rates \cite{BaalrudDaligault2013prl,BaalrudDaligault2014pop}, but the inaccuracy beyond this threshold is not as severe for these coefficients.  



\section{Summary}

We have carried out a detailed study of the calculation of the shear
viscosity coefficient of one-component plasmas with equilibrium MD simulation in order to
independently validate the non-equilibrium MD results of
\cite{DonkoHartmann2008} for $\kappa>0$ and the equilibrium MD
simulations of \cite{bast:05}.
We have presented a convergence study of the Green-Kubo relation to
determine optimal simulation parameters and, in turn, produce
accurate viscosity coefficients.
Finally, we have compared the accurate data to various sophisticated
theoretical predictions.


\acknowledgments
This work was carried out under the auspices of the National Nuclear Security Administration of the U.S. Department of Energy (DOE) at Los Alamos National Laboratory under Contract No. DE-AC52-06NA25396.
The work of J.D. and K.\O.R. was supported by the DOE Office of Fusion Sciences.
The work of S.D.B was supported in part by the University of Iowa and in part by Los Alamos National Laboratory.

\appendix

\section{Kinetic-kinetic term}

The inital value of the kinetic-kinetic contribution to the shear stress correlation function is
\ben
\big\langle \sigma_{xy}^{\rm kin}(0) \sigma_{xy}^{\rm kin}(0) \big\rangle_{eq}=\lim_{t\to\infty}{\bar{J}_{xy}^{kin}(t)}
\een
where
\begin{widetext}
\ben
\bar{J}_{xy}^{kin}(t)&=&
\frac{1}{t}\int_0^{t}{\sum_{i=1}^N{m v_{x,i}(s)^2}\sum_{j=1}{mv_{y,j}(s)^2}ds  }+\frac{1}{t}\int_0^{t}{ds
\sum_{i\neq j=1}^N{m^2 v_{x,i}(s)v_{x,j}(s)v_{y,i}(s)v_{y,j}(s)}ds}\\
&=&\left[\frac{1}{t}\int_0^{t}{\sum_{i=1}^N{m v_{x,i}(s)^2}ds}\right]\left[\frac{1}{t}\int_0^{t}{\sum_{i=1}^N{m v_{y,i}(s)^2}ds}\right]+\text{cross terms}(t)
\een
and
\be
\text{cross terms}(t)&=&\frac{1}{t}\int_0^{t}{\left[\sum_{i=1}^N{m v_{x,i}(s)^2}-\frac{1}{t}\int_0^{t}{\sum_{i=1}^N{m v_{x,i}(s)^2}ds}\right]}\times\left[\sum_{i=1}^N{mv_{y,i}(s)^2}-\frac{1}{t}\int_0^{t}{\sum_{i=1}^N{m v_{y,i}(s)^2}ds}\right]ds\nn\\
&&+\frac{1}{t}\int_0^{t}{\sum_{i\neq j=1}^N{m^2 v_{x,i}(s)v_{x,j}(s)v_{y,i}(s)v_{y,j}(s)}ds}\label{cross_term}
\ee
\end{widetext}
In the limit $t\to\infty$, 
\ben
\left[\frac{1}{t}\int_0^{t}{\sum_{i=1}^N{m v_{x,i}(s)^2}ds}\right]\left[\frac{1}{t}\int_0^{t}{\sum_{i=1}^N{m v_{y,i}(s)^2}ds}\right]\\
=(Nk_BT)(Nk_BT)
\een
and
\ben
\text{cross terms}(t)= 0
\een


\begin{references}
\bibitem{BausHansen1980} M. Baus and J.-P. Hansen, {\it Phys. Rep.} {\bf 59}, 1 (1980).
\bibitem{SaigoHamaguchi2002} T. Saigo and S. Hamaguchi, {\it Phys. Plasmas} {\bf 9}, 1210 (2002).
\bibitem{VieillefosseHansen1975} P. Vieillefosse and J.P. Hansen, {\it Phys. Rev. A} {\bf 12}, 1106 (1975).
\bibitem{Daligault2006} J. Daligault, {\it Phys. Rev. Lett.} {\bf 96},
  065003 (2006). The MD results for the viscosity coefficients
  published in this paper are incorrect because of an unintentional mistake of the author in
  implementing the formula for the viscosity in his code.
\bibitem{DonkoHartmann2008} Z. Donk{\'o} and P. Hartmann, {\it
    Phys. Rev. E} {\bf 78}, 026408 (2008).
\bibitem{bast:05} S.\ Bastea, {\it Phys.\ Rev.\ E} {\bf 71}, 056405
  (2005).
\bibitem{bern:78} B.\ Bernu and P.\ Vieillefosse, {\it Phys.\ Rev.\ A} {\bf 18}, 2345 (1978). 
\bibitem{wall:78} J.\ Wallenborn and M.\ Baus, {\it Phys.\ Rev.\ A} {\bf 18}, 1737 (1978). 
\bibitem{SalinCaillol2003} G. Salin and J.-M. Caillol, {\it Phys. Plasmas} {\bf 10}, 1220 (2003).
\bibitem{HockneyEastwood} R. Hockney and J. Eastwood, {\it Computer
    Simulation using Particles} (IOP Publishing, 1988).
\bibitem{FrenkelSmit} D. Frenkel and B. Smit, {\it Understanding
    Molecular Dynamics} (Academic Press, 2002).
\bibitem{HansenMcDonald} J.P. Hansen and I.R. McDonald, {\it theory of Simple Liquids} (Academic, London, 1986).
\bibitem{noteDaligault} J. Daligault, unpublished.
\bibitem{ZwanzigAilawadi1969} R. Zwanzig and N.K. Ailawadi, {\it Phys. Rev.} {\bf 182}, 280 (1969).
\bibitem{Bitsanis1987} I. Bitsanis, M. Tirrell and H. Ted Davis, {\it Phys. Rev. A} {\bf 36}, 958 (1987).
\bibitem{Spitzerbook} L. Spitzer, Jr., {\it Physics of Fully Ionized Gases, 2nd Ed.} (Interscience, New York, 1962).
\bibitem{DimonteDaligault2008} G. Dimonte and J. Daligault, Phys. Rev. Lett. {\bf 101}, 135001 (2008).
\bibitem{BaalrudDaligault2013prl} S.D. Baalrud and J. Daligault, {\it
    Phys.\ Rev.\ Lett.} {\bf 110}, 235001 (2013).
\bibitem{Daligault2012} J. Daligault, Phys. Rev. Lett. 108, 225004 (2012).
\bibitem{IchimaruVol1} S. Ichimaru, {\it Statistical Plasma Physics, Vol. I: Basic Principles}, Addison-Wesley Publ. Company (1992).
\bibitem{tana:86} S.\ Tanaka and S.\ Ichimaru, {\it Phys.\ Rev.\ A} {\bf 34}, 4163 (1986). 
\bibitem{Braun1967} E. Braun, {\it Phys. of Plasmas} {\bf 10}, 731
  (1967).
\bibitem{BaalrudDaligault2014pop} S.D. Baalrud and J. Daligault, {\it
    Phys. Plasmas} {\bf 21}, 055707 (2014).
\bibitem{saig:02} T.\ Saigo and S.\ Hamaguchi, {\it Phys.\ Plasmas} {\bf 9}, 1210 (2002). 
\end{references}
\end{document}